\documentclass[structabstract]{aa}  
\usepackage{graphicx}
\usepackage{longtable}
\usepackage{lscape}
\usepackage{cite}
\usepackage{natbib} 
\usepackage{txfonts}
\def\gsim{\mathrel{\hbox{\rlap{\hbox{\lower4pt\hbox{$\sim$}}}\hbox{$>$}}}}
\def\lsim{\mathrel{\hbox{\rlap{\hbox{\lower4pt\hbox{$\sim$}}}\hbox{$<$}}}}

\begin{document}
\title{Dust in active galactic nuclei}
   \subtitle{Mid-infrared T-ReCS/Gemini spectra using the new RedCan pipeline}
   \author{O. Gonz\'alez-Mart\'in\inst{1,2}\thanks{Juan de la Cierva Fellow (\email{omairagm@iac.es})},
   J.M. Rodr\'iguez-Espinosa\inst{1,2}, T. D\'iaz-Santos\inst{3}, C.Packham\inst{4},\\ A. Alonso-Herrero\inst{5,6}, 
   P. Esquej\inst{6,7,8}, C. Ramos Almeida\inst{1,2}, R. Mason\inst{9}, \and C. Telesco\inst{10} 
	  }

   \institute{Instituto de Astrof\'isica de Canarias (IAC), C/V\'ia L\'actea, s/n, E-38205 La Laguna, Spain \and 
   	Departamento de Astrof\'isica, Universidad de La Laguna (ULL), E-38205 La Laguna, Spain \and 
      Spitzer Science Center, California Institute of Technology, 1200 East California Boulevard, Pasadena, CA 91125, USA \and
      Dept. of Physics \& Astronomy, University of Texas San Antonio, San Antonio, TX. 78249, USA \and
      Augusto Gonzalez Linares Senior Research Fellow \and
      Instituto de F\'isica de Cantabria, CSIC-Universidad de Cantabria, 39005 Santander, Spain \and 
      Departamento de F\'isica Moderna, Universidad de Cantabria, Avda. de Los Castros s/n, 39005 Santander, Spain \and
      Centro de Astrobiolog\'a, INTA-CSIC, E-28850 Madrid, Spain \and 
      Gemini Observatory, Northern Operations Center, 670 N. A'ohoku Place, Hilo, HI 96720, USA \and
      Department of Astronomy, University of Florida, Gainesville, Florida 32611, USA 
    }
   \date{Received ?? 2012; accepted ??, 2012}
\authorrunning{O.Gonz\'alez-Mart\'in et al.}
\titlerunning{Dust in AGN}
\abstract  
   {The unified model of active galactic nuclei (AGN) claims that the properties of AGN
   depend on the viewing angle of the observer with respect to a toroidal distribution of
   dust surrounding the nucleus. Both the mid-infrared (MIR) attenuation and
   continuum luminosity are expected to be related to dust associated with the torus. 
   Therefore, isolating the nuclear component is essential to study the MIR
   emission of AGN. }
   {This work is aimed to study the MIR emission of AGN with the highest spatial
   resolution available to date, isolating its contribution from extended emission. We
   would like to address three fundamental questions: (1) how important is the AGN
   contribution to the MIR spectrum? (2) does dust attenuation arise mainly from the
   central dusty torus associated to the AGN? and (3) how does spatial resolution affect
   these issues?}  
   {We have compiled all the T-ReCS spectra (Gemini observatory) available in the N-band
   for 22 AGN: 5 Type-1 and 17 Type-2 AGN. The high angular resolution of the T-ReCS
   spectra allows us to probe physical regions of 57 pc (median). We have used a novel
   pipeline called RedCan capable of producing flux- and wavelength-calibrated spectra for
   the CanariCam (GTC) and T-ReCS (Gemini) instruments. We have measured the fine-structure
   [S~IV] at 10.5$\rm{\mu m}$ and the PAH at 11.3$\rm{\mu m}$ line strengths together with
   the silicate absorption/emission features. We have also compiled \emph{Spitzer}/IRS
   spectra to understand how spatial resolution influences the results. We
   have complemented our sample with the results of 19 VISIR/VLT spectra (Paranal
   observatory) published by H{\"o}nig et al. (2010a) and 20 nearby highly obscured AGN
   ($\rm{N_H> 1.5\times 10^{24}cm^{-2}}$) \emph{Spitzer} spectra published by Goulding et
   al. (2012).}
   {The 11.3$\rm{\mu m}$ PAH feature is only clearly detected in the nuclear spectra of two
	AGN, while it is more common in the \emph{Spitzer} data. For those two objects the AGN
	emission in NGC\,7130 accounts for more than 80\% of the MIR continuum at 12$\rm{\mu m}$
	while in the case of NGC\,1808 the AGN is not dominating the MIR emission. This is
	confirmed by the correlation between the MIR and X-ray continuum luminosities. The [S~IV]
	emission line at 10.5$\mu m$, which is believed to originate in the narrow line region, is
	detected in most AGN. We have found an enhancement of the optical depth at 9.7$\mu
	m$ ($\tau_{9.7}$) in the high-angular resolution data for higher values of $\rm{N_{H}}$.
	Clumpy torus models reproduce the observed values \emph{only} if the host-galaxy
	properties are taken into account.}    
 

\maketitle


 \begin{table*}
  \begin{center}
 \scriptsize{
 \begin{tabular}{l r c c c c c c c c}
\hline\hline  	       
Num &     Name   &  D(Mpc) &  Slit width(arcsec)/Size(pc) & Type  &  Optical features & log(NH)  & log(L(2-10keV)) & X-ray Ref & $b/a$ \\
     (1)	  &  (2)   &   (3)	 & (4)    &  (5)  &   (6) & (7)& (8) & (9)  & (10)  \\ \hline		       
		   &  	     &	   	 &	  &       &	  &    &   &    &  \\
   1   & 	  NGC\,1365 & 17.9    &       0.35/30  & S1.5  &       & 23.6   & 42.5  & (1,11)  &  0.5 \\ 
   2   & 	  NGC\,1386 & 16.2    &       0.31/24  & S2    &E      &$>$25.0 & 42.6  & (2,3)  &  0.4 \\
   3   & 	  NGC\,1808 & 11.5    &       0.35/19  & S2    &       & 22.9   & 40.4  & (3,12)  &  0.6    \\
   4   & 	  NGC\,3081 & 32.5    &       0.65/102  & S2    &       & 23.9   & 42.5  & (2)  &  0.8    \\
   5   & 	  NGC\,3094 & 44.7    &      0.35/76  & S2    &       &        &       &      &  0.7    \\
   6   & 	  NGC\,3281 & 45.7    &      0.35/77  & S2    &D,E    & 24.3   & 43.4  & (1,12)  &  0.4    \\
   7   & 	  NGC\,4418 & 31.1    &      0.35/53   & S2    &E      &$>$25.0 & 41.2  & (4)  &  0.4    \\
   8   & 	  NGC\,4945 &  4.0    &      0.65/13   & S2    &D,E,M  & 24.7   & 42.3  & (5,12)  &  0.4    \\
   9   &   Centaurus\,A &  3.7    &      0.65/12   & S2    &D,E    & 23.7   & 41.9  & (1,13)  &  0.4   \\
   10  & 	  NGC\,5135 & 58.6    &     0.70/199   & S2    &       &$>$25.0 & 43.1  & (2)  &  0.7    \\
   11  & 	  Circinus  & 4.2     &       0.35/7   &  S2   &M      & 24.6   & 42.6  & (6,2) &   0.5    \\
   12  & 	  NGC\,5506 & 28.8    &      0.35/49   & S1.9  &D,E    & 22.5   & 43.0  & (1,3) &   0.2    \\
   13  & 	  NGC\,5643 & 16.9    &      0.35/29   & S2    &       & 23.8   & 42.6  & (7,12) &   0.9    \\
   14  & 	  NGC\,5728 & 29.1    &      0.35/49   & S1.9  &       & 23.6   & 43.3  & (6) &   0.6    \\
   15  & 	  IC\,4518W & 69.6    &     0.70/236   & S2    &E,M    & 24.3   & 42.6  & (8,13) &   0.1    \\
   16  &     ESO\,103$-$G35 & 56.9    &      0.35/96  & S2    &E      & 23.2   & 43.4  & (1,12) & 0.4    \\
   17  & 	   IC\,5063 & 48.6    &     0.65/153   & S2    &D    & 23.3   & 43.3  & (1,12)  &   0.7   \\  
   18  & 	  NGC\,7130 & 69.2    &     0.70/235   & L2    &M      &$>$25.0 & 43.1  & (10,2) &  0.9    \\
   19  & 	  NGC\,7172 & 33.9    &      0.35/57   & S2    &D    & 22.9   & 42.8  & (1,3)  &  0.6    \\  
   20  & 	    3C\,445 &239.3    &     0.35/406  & S1.5  &       & 23.5   & 43.9  & (3)  &   0.9    \\
   21  & 	  NGC\,7479 & 33.8    &        0.35/57 & L1.9  &D,M    & 24.3   & 42.0  & (3)  &   0.8   \\
   22  & 	  NGC\,7582 & 20.6    &       0.70/70  & S2    &D,E    & 23.0   & 42.6  & (2)  &   0.4    \\ \hline
\hline
\end{tabular}
\caption{Sample properties. The size of the region observed by T-ReCS is computed using the slit width.
Optical features: (D) Dust features across the nuclear region; (E) Edge-on galaxies; (M) Merger. The references for the X-ray results are: 
(1) \citet{Tueller08},
(2) \citet{Marinucci12},
(3) \citet{Brightman11},
(4) \citet{Maiolino03},
(5) \citet{Itoh08},
(6) \citet{Goulding12},
(7) \citet{Guainazzi04},
(8) \citet{Pereira-Santaella11},
(9) \citet{Moran96},  
(10) \citet{Mullaney11},
(11) \citet{Risaliti09},
(12) \citet{Comastri04},
(13) \citet{delaRosa11}, and 
\label{tab:sample}  }
}
\end{center}
\end{table*}

\section{Introduction}\label{sec:intro}

The radiation from active galactic nuclei (AGN) is due to accretion onto a super-massive
black hole (SMBH). The optical spectra of AGN have been classified into two main classes
according to the existence (Type-1) or not (Type-2) of broad permitted lines
(FWHM$>2000~km~s^{-1}$). The \emph{so-called} unified model proposes that both types of
AGN are essentially the same objects viewed at different angles
\citep{Antonucci93,Urry95}. An optically-thick dusty torus surrounding the central source
would be responsible for blocking the region where these broad emission lines are produced
(the \emph{so-called} broad line region, BLR). Therefore, Type-2 AGN are essentially
Type-1 AGN blocked by the dusty torus along the line of sight (LOS) to the observer. More
elaborated models suggest that the same observational constraints can also be explained
with a torus made of discrete dense molecular clouds following a certain radial
distribution \citep{Honig06,Schartmann08,Elitzur08,Nenkova08a,Nenkova08b,Honig10b}.

Mid-infrared (MIR) spectroscopy is a powerful tool to examine the nature of AGN, as well
as the contribution of star-formation activity. \citet{Weedman05} found a variety of
spectral features displayed in the MIR spectra of eight classical AGN with \emph{Spitzer}
data \citep[see also][]{RodriguezEspinosa87,RodriguezEspinosa97}. Large samples of AGN
observed with \emph{Spitzer} have been studied in detail to quantify the properties of AGN
at MIR \citep[e.g.][]{Sturm06,Deo07,Hao07}.

Among the most noteworthy results found at MIR wavelengths, \citet{Wu09} found that
the silicate features at 9.7$\rm{\mu m}$ and 18$\rm{\mu m}$ in Type-1 Seyfert galaxies are
rather weak, while Type-2 Seyfert galaxies are likely to correspond to strong silicate absorptions.
Furthermore, the deepest silicate absorption features show the highest values for the
hydrogen column density ($\rm{N_{H}}$) in X-rays \citep[][]{Shi06}. The silicate emission
feature in Type-1 AGN is much weaker than expected from smooth dusty tori. The clumpiness
of the dusty torus explains better the overall suppression of this feature
\citep{Nenkova02,Nenkova08a}. However, clumpy models do not predict strong silicate
absorption features because illuminated and shadowed clouds contribute to the observed
spectrum. In fact, the silicate feature depth typically seen in some Compton-thick AGN
($\rm{N_{H}>1.5\times 10^{24}cm^{-2}}$) observed with \emph{Spitzer} is similar to that
observed in Type-1 AGN \citep{Goulding12}. However, the nature of these silicate feature
is not clear since the host-galaxy could be another contributor to the dust attenuation
seen in the MIR \citep[see also][]{Alonso-Herrero11,Deo07}.

Another prominent feature noticeable in the MIR is the polycyclic aromatic hydrocarbon
(PAH) emission. This is usually prominent in starburst galaxies, with a good correlation
between the strength of the PAHs and the IR luminosity in starbursts \citep{Brandl06}.
This correlation appears however to be weak or absent in AGN \citep{Weedman05}. 
PAHs form in the photodissociation regions (PDRs) in star forming clouds, but these PDRs
can easily be destroyed by the intense radiation field from the AGN, even if it is a
vigorous starburst \citep{Genzel98,Laurent00}. \citet{Goulding12}, using low resolution
\emph{Spitzer} spectra, found that Compton-thick AGN show strong PAH features, suggesting
that the PDRs can survive the AGN radiation field if properly shielded.
\citet{Diamond-Stanic12} found a relation between the accretion rate of AGN and the SFR
derived from PAHs. They suggest that the gas fueling the SMBH is also used for the
star-formation.

However, the relatively low-spatial resolution of \emph{Spitzer}/IRS spectra ($\sim 3$
arcsec) include a non-negligible contribution of the host galaxy and/or of circumnuclear
star formation. Thus, a good isolation of the torus emission is needed to understand the
pure AGN contribution to the MIR spectrum. The best isolation possible to date can
be obtained using IR interferometry \citep[e.g.][]{Tristram07, Tristram09}. However, this
can only be done for a few bright AGN. As for the rest of AGN the medium- to large-size
ground-based telescopes offer an opportunity to better isolate the contribution of the AGN
within a few tens of parsecs. Some individual works have been presented showing the
importance of spatial resolution to isolate the AGN from the host galaxy \citep[e.g.
NGC\,1068, NGC\,1365, NGC\,3281 and the Circinus galaxy;][respectively]{Mason06,
Alonso-Herrero12, Sales11, Roche06}. A large sample was presented by \citet{Honig10a},
including 19 AGN observed with the VLT/VISIR MIR spectrograph with spatial resolution
better than $\sim$100 pc. They found moderately deep silicate absorption features.
However, this sample is free of objects with high inclination angles or dust lanes, and
include only three Compton-thick AGN. The MIR luminosity seems to be quite isotropic,
correlating with other AGN luminosity tracers \citep[see
also][]{Ramos-Almeida07,Levenson09,Gandhi09,Mason12}. 

In this paper, we compile all the AGN observed by T-ReCS at the Gemini observatory. The
sample comprises 22 AGN, including 9 Compton-thick sources. Our main aims are: (1) to
characterize the AGN pure emission at MIR frequencies; (2) to understand the nature of the
attenuation seen at MIR frequencies as traced by the optical depth at 9.7$\mu m$; (3) to
compare T-ReCS and \emph{Spitzer} spectra to understand the limitations introduced by the
lower spatial resolution in \emph{Spitzer} spectra.

For the data processing and analysis we have used a newly developed pipeline called
RedCan\footnote{The CanariCam data reduction pipeline.}. This pipeline is able to produce
flux-calibrated images and 1D spectra for both the T-ReCS \citep{Telesco98,DeBuizer05} and
the CanariCam\footnote{CanariCam is a mid-infrared (7.5 - 25$\mu m$) imager with
spectroscopic, coronagraphic, and polarimetric capabilities, which is mounted at one of
the Nasmyth foci of the GTC telescope at El Roque observatory.} \citep{Telesco03}
instruments without user interaction. This pipeline has been produced to facilitate
a better exploitation of the CanariCam data and can be freely accessed (see Section
\ref{sec:reduction}). 

\begin{figure*}[!t] 

\begin{center}

\includegraphics[width=2.0\columnwidth]{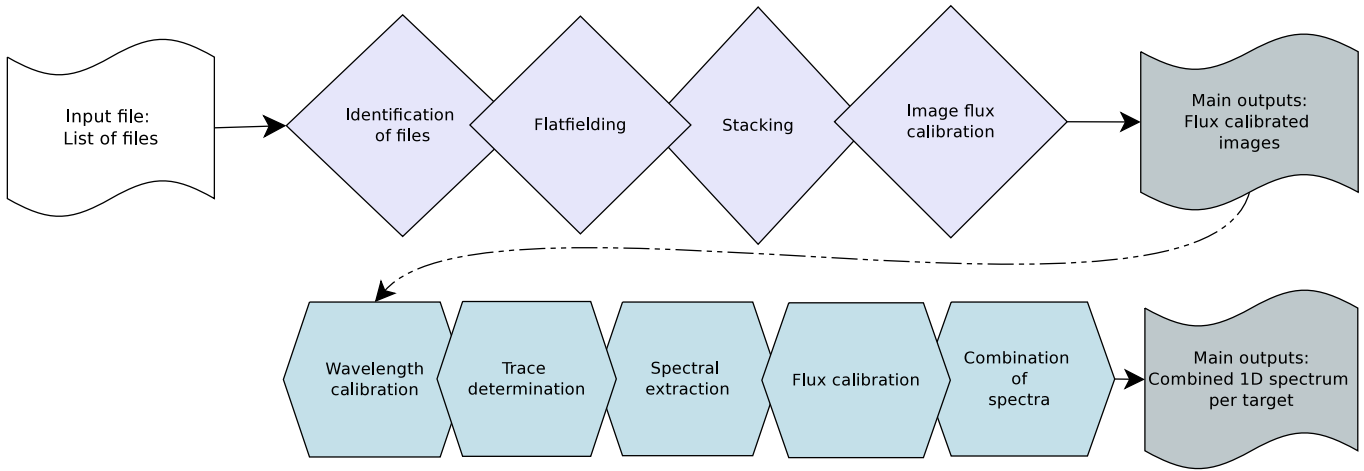}

\caption{Flowchart of the RedCan procedure. Diamonds are the steps of the imaging
procedure while polygons indicate the steps of the spectroscopic procedure. The empty flag
is the input level and the dark gray flags indicate the outputs levels. The dashed arrow
is set to remark that RedCan continues from imaging to spectroscopic procedures only when
spectra are included in the initial dataset. }

\label{fig:RedCan-scheme} 

\end{center} 

\end{figure*}

\section{Sample selection}\label{sec:sample}

We have searched for all the T-ReCS (Gemini South observatory) AGN spectra available in
the Gemini public
archive\footnote{http://www3.cadc-ccda.hia-iha.nrc-cnrc.gc.ca/cadcbin/gsa/wdbi.cgi/gsa/gsa
\_science/form} until June 2012. The sample comprises 22 AGN observed as part of $\sim$10
observational projects. All of them are observed in the N-band. Note that, since T-ReCS is
not expected to be available after 2013A this sample includes most of the AGN ever
to be observed with this instrument\footnote{Except for three objects that are not
public yet.}.

Table~\ref{tab:sample} shows the main observational details of the sample. Seventeen objects are
Type-2 and five are Type-1 AGN. The slit width used for the spectroscopy results in
a spatial resolution between $\sim 7-236$ pc for all the objects except for 3C\,445
(406~pc), which is by far the most distant object in the sample. This is the largest
sample of high-spatial resolution spectra ever put together, especially when including the
19 AGN (VISIR/VLT) reported by \citet{Honig10a}. Our sample includes 9 Compton-thick
AGN (see $\rm{N_{H}}$ in Table \ref{tab:sample}), making it complementary to the AGN
sample in \citet{Honig10a} (where only 3 Compton-thick AGN are included). Note, however,
that our sample is unbiased though not complete in any sense since the observations
come from several different proposals with differing goals. It is worth noticing that this
sample appears to contain a high fraction of sources with large silicate absorption
features (see Section \ref{sec:silicate}).

\section{Data reduction: the CanariCam data reduction pipeline (RedCan)}\label{sec:reduction}

Data reduction pipelines are an important aspect of modern astronomy. Indeed large
telescope instruments are becoming fairly complex, hence the prompt availability of
science data is crucial for securing fast scientific turnout from today's large and
expensive facilities. Unfortunately, this is not always the case, even in successful
facilities. The aim of our group is to produce a new and advanced pipeline for MIR
imaging and spectroscopic data. In the forthcoming years large amounts of GTC/CanariCam
data will be acquired by our group. This includes data from both guaranteed CanariCam time
($\sim$100 hours, PI C. Telesco) and the ESO/GTC proposal led by A. Alonso-Herrero
($\sim$180 hours, ESO/GTC 182.B.2005). It is therefore important to have a data reduction
pipeline that facilitates the use of the data as soon as possible after being acquired at
the telescope. This is the purpose of RedCan, the Canaricam data reduction pipeline we
have developed. RedCan is a well tested reduction and calibration procedure to ensure reliable
data products from CanariCam.

Thanks to the similarities between the T-ReCS and CanariCam instruments, RedCan can be
used for data from both instruments in both imaging and spectroscopy modes (Q and N bands
in medium resolution).

RedCan has been written combining the Gemini IRAF package\footnote{The released version of
the Gemini IRAF package is an external package layered upon IRAF and is available to users
and other interested parties (
http://www.gemini.edu/sciops/data-and-results/processing-software}) within PyRAF,
together with ad hoc IDL routines developed for this purpose. CSHELL language is used
for the main routine of the pipeline to connect IDL and Python routines. \footnote{The
RedCan package can be retrieved here: {\scriptsize
https://www.dropbox.com/sh/w4z8buo2ewrhhvj/Kk1f2RYVe8/RedCan.tar.gz}. Note that the
pipeline is of free access although it will not be supported. A manual is also available
that includes documentation on the software requirements and installation procedures.}.
Fig.~\ref{fig:RedCan-scheme} shows a chart flow of the reduction steps done by RedCan to
produce the final reduced data. RedCan is based on the procedure described by
\citet[][]{Diaz-Santos10} \citep[see also][]{Diaz-Santos08} and the main input file is
an ASCII file with a list of observations. A full description of the pipeline procedure is
given below.

\subsection{Identification of files}

A full data set from a target observation consists of: (1) telluric standard observations,
(2) target observation, and (3) acquisitions of both standard and target images. Standards
and targets can be in imaging and/or spectroscopic modes. Moreover, the acquisition images in
the case of spectroscopic data can be either imaging or imaging throughout the slit.

RedCan uses keywords in the fits headers to identify the type of each observing block,
which are used accordingly throughout the pipeline. In particular the keyword {\sc obsclass}
together with the instrumental configuration keywords (i.e. {\sc slit, grating, filter1,
filter2}, etc.) and observational keywords (i.e. {\sc ontime, ra, dec, ut, date}) are used
to identify the following information:

\begin{itemize}

\item Acquisition image associated to a target/standard. RedCan sorts the files according
to the observed time and selects as acquisition images those immediately before the
target/standard. Acquisition images of the objects through the slit are used to compute
slit losses for the spectroscopic mode (see Section~\ref{sec:specfluxcal}). These images
are also flux calibrated during the process so they can be used for science purposes.

\item Standard star image/spectrum associated to a target. RedCan selects within the same
date of observation the standard image/spectrum observed with the same configuration used
for the target (i.e. same filter and grating configuration with the same slit size). Note
that if more than one observation of standard stars fulfill this criterion, the one closer
in time is chosen. This association is used to flux calibrate both images and spectra (see
Sections \ref{sec:imagefluxcal} and \ref{sec:specfluxcal}).

\end{itemize}

\subsection{Flat fielding}

The RedCan pipeline allows to subtract a master flat computed as an average of the
observations of the sky for each chop integration. This master-flat is computed using the
tasks {\sc tprepare} and {\sc mistack} within the midir/IRAF package while the flat
subtraction is implemented in IDL. \citet{Diaz-Santos08} concluded that flat-fielding can
improve the quality of the final images. However, the flat-fielding procedure introduces noise
into the final image (see e.g. the Gemini web page\footnote{
http://www.gemini.edu/?q=node/10139\#flat}). We have detected that using flat-fielding in
the spectroscopic mode can lead to follow a wrong trace of the spectra due to the
inclusion of patterns at the borders of the array. This step is included in RedCan as an
option for the user, although the default is set to skip it. We have skipped this option
in the data presented in this paper.

\subsection{Stacking}\label{sec:stacking}

Accurate cancellation of the sky and telescope background is most readily achieved through
the technique of chopping and nodding. The sky is very bright and highly variable at
MIR frequencies when observed from the earth. To get rid of the sky, the chop technique
observes frequently a nearby sky position (off-source chop) moving the secondary mirror of
the telescope to subtract from the target position (on-source chop). However, this
chopping technique leaves a background resulting from the different paths followed by the
light depending on whether the telescope is pointing to the on-source or the off-source
positions. The nodding technique places the source at the nod positions moving the actual
telescope axes to remove this residual. Thus, the chop-nod technique allows the
subtraction of the sky background as well as the background from the telescope itself.
This technique also removes the array 1/f noise. The data are then stored automatically in
save-sets. Prior to stacking, RedCan automatically rejects chop-nod pairs of images
affected by noise patterns or glitches due to instrumental or background issues during the
observations. This rejection is done by default excluding any chop-nod pair above or below
5-$\sigma$ of the mean value of the subtracted chop-nod pair of images (task {\sc
tbackground} within Gemini/midir). However, this can be executed manually and the
threshold level can be changed if desired. Both the on-source and the off-source chop
images are averaged together (task {\sc miprepare}) and each nod pair combined (task {\sc
mistack}). The resulting image shows certain types of pattern noise as horizontal or
vertical stripes. Those that are constant along the image are removed with the task {\sc
miclean} within the gemini/midir package. Finally each image is divided by the frame time
so the final image is in units of ``adu/s''. 

The final product is a single data cube per observation with three planes: on
source, off source, and on$-$off source (i.e. background subtracted) images. This
procedure is done for every image (i.e. acquisitions, standards, and targets).  Note that
the second plane is needed for wavelength calibration in the spectroscopic mode (see
Section \ref{sec:wlcal}).

 \begin{table}
 \begin{center}
 \scriptsize{
 \begin{tabular}{l l r r r r r}
\hline\hline  	       
Name   & ObsID  &   Expt.  &   Date &  P.A.  & STD*  &  FWHM  \\
		   &  	     &	   	    &	       &	&	&   (8.7$\mu m$) \\
     (1)	  &  (2)   &   (3)	 & (4)    &  (5) & (6)  & (7)	   \\ \hline		      
		   &  	     &	   	    &	       &	&	&    \\
  NGC\,1365 &	  0061 &    624  &    090829   &     	  15 &     26967  & 0.14 \\
   NGC\,1386 &	  0210,12 &   2x 606    &  060929   &     	   0 &    23319  & 0.29 \\
  NGC\,1808 &	  0076 &    981   &   090829  &	    	  45 &     26967  & 0.20 \\
  NGC\,3081 &	  0094 &    606   &   060125   &     	   0 &    81797  & 0.20 \\
          &	  0113 &    606    &  060304   &     	 350 &    93813   &  0.29  \\
  NGC\,3094 &	  0075,76 &    2x309   &   040506   &     	   0 &    81797   & 1.10   \\
  NGC\,3281 &	  0120 &    981   &   090328   &     	 315 &  HR3438*   &  0.17 \\
          &	  0033,35 &    2x981   &  090406   &     	 315 &   146791   &  0.20  \\
  NGC\,4418 &	  0041 &    309   &   040507   &     	  40 &   128620   &  0.25 \\
          &	  0046,47 &    2x309   &   040508   &     	  40 &   128620   &  0.13  \\
          &	  0025-28 &    4x327   &   080628   &     	  30 &    99998   &  0.30 \\
  NGC\,4945 &	  0055,56 &    2x303    &  060417   &     	  45 &    110458  &  0.26 \\
Centaurus\,A &	  0198 &    517    &  040131   &     	   0  &   123139   & 0.22  \\
  NGC\,5135 &	  0098 &    633    &  060306   &     	  30  &   101666  &  0.19  \\
          &	  0106,07 &    2x633    &  060310   &     	  30   &   101666  &  0.16 \\
 CIRCINUS &	  0084,85 &    2x309    &  040506   &     	 100 &    81797   &  1.10 \\
          &	  0095-97 &    3x309    &  040506   &     	  10 &    81797   &  1.10 \\
  NGC\,5506 &	  0107,08 &    2x309    &  040506   &     	   0 &    81797  &  1.10 \\
  NGC\,5643 &	  0057-60 &    4x303    &  070605   &     	  80 &   123139   &  0.21 \\
  		 &	  0057-59 &  3x327 &    080615   &	  80 &   123139   &  0.50  \\
  		 &	  0060-62 &  6x327 &    080615   &	  80 &    139127   &  0.40  \\
  NGC\,5728 &	  0046,47 &    2x921   &   090406   &     	 215 &   146791  &  0.18  \\
  IC\,4518W &	  0081-83 &    3x633    &  060417   &     	   5  &   123123  &  0.22  \\
ESO\,103$-$G35 &	  0065-67 &    3x318   &   080616   &     	  30 &   156277  &  0.24 \\
                  &	  0068-70 &    3x318   &  080616   &     	  30 &   171759   &  0.51  \\
   IC\,5063 &	  0078 &    606    &  05-05-21   &     	 303 &    81797  &  0.20  \\
   		 &	  0046 & 606 &      050707   &	 303 &    81797  &  0.20  \\
   		 &	  0047 & 303 &     050707   &	 303 &    81797  &  0.20  \\
   		 &	  0048 & 302 &     050707   &	 303 &    81797  &  0.20  \\
   		 &	  0092 & 302 &     050709   &	 303 &    81797  &  0.20  \\
  NGC\,7130 &	  0061 &    633  &    050918   &     	 348 &   190056  &  0.30  \\
               &	  0062 &    633  &    050918   &     	 348 &  219784  &  0.35  \\
          	&	  0111-113 &  3x633 &    060704   &	 348 &   190056  &  0.14 \\
          	&	  0057 &  633 &    060829   &	 348 &   190056  &  0.24  \\
          	&	  0107,08 &  2x633 &    060916   &	 348 &   190056  &  0.19  \\
          	&	  0097,98 &  2x633 &    060925   &	 348  &   190056  &  0.25  \\
          	&	  0099 &  633 &    060925   &	 348  &   219784  &  0.29 \\
  NGC\,7172 &	  0120-22 &    3x309  &    040506   &     	  60  &    81797  & 1.10 \\
  		 &	  0077-80 &  4x327 &    080615   &	  90 &   200914  & 0.29  \\
  		 &	  0041,42 &  2x318 &    080625   &	  90 &   200914  & 0.34  \\
  		 &	  0063-66 &  4x327 &    080627   &	  90 &   200914  & 0.25 \\
    3C\,445 &	  0098,99 &    2x921   &   100606    &    	 300 &   206445  &  0.24  \\
  NGC\,7479 &	  0055-57 &    3x327   &   080629   &     	  10 &   206445  &  0.41  \\
                   &	  0058-60 &    3x327   &   080629   &     	  10 &     1255  &  0.23  \\
  NGC\,7582 &	  0174,75 &    2x606   &   050718   &     	   0  & BS8556*  &  0.20  \\
\hline
\end{tabular}
\caption{Observational details of the T-ReCS spectra. *STD: Standard star used to perform flux-calibration.
All the standard stars are named with their HD sequence number except those marked with a star.  
The ObsID (Col. 2) includes the last 4 digits of the name of the observing file. The first 8 digits are the date
of the observation (i.e. ``S"+ObsID+``S20"+YYMMDD).  }
\label{tab:observations}  
}
\end{center}
\end{table}

\subsection{Imaging flux calibration} \label{sec:imagefluxcal}

Images resulting from the stacking process described above will now be flux calibrated.
This flux calibration is performed for both acquisition and target images using associated
telluric standards stars.

RedCan contains the catalog of synthetic flux-calibrated spectra of standards stars
published by \citet{Cohen99}. An identification of the standard star is done using the
coordinates given in the header of the observation. When the standard star is not part of
the catalog of Cohen, a reference standard star is used and a warning is shown in the
screen and recorded in the output tables and log of the reduction process. Then, the
synthetic spectrum of the star is convolved with the telescope throughput and filter
bandpass. Aperture photometry of the standard star is performed and a new keyword called
{\sc fcalstd} is placed in the standard star header which provides the ratio between the
number of counts per second and the real flux of the star. Then aperture photometry
techniques are applied to the target images and a table with the final fluxes for the set
of images is recorded. Note that aperture radii of the standard star and target are fixed
to a radius of 2.7 and 0.9 arcsec respectively. These radii include more than 95\% of
the flux at the diffraction limit in both T-ReCS (FWHM$\sim$0.35-0.37 arcsec
\footnote{http://www.gemini.edu/?q=node/10582\#trecspsf}) and CanariCam data
(FWHM$\sim$0.26-0.42 arcsec
\footnote{http://www.gtc.iac.es/en/pages/instrumentation/canaricam/data-commissioning.php
\#Image\_Quality}) and above 90\% with good seeing conditions (optical seeing below 1 arcsec). The
background is extracted using annuli with internal and external radii of [4.5, 5.3] and
[1.8, 2.25] arcsec, respectively. Finally, flux calibrated images are also produced for
target and standard stars using the conversion factor stored in {\sc fcalstd}.

This method is flux accurate to 10\% \citep{Diaz-Santos08}. However, we note that this
aperture photometry is accurate enough for point-like sources while a more sophisticated
methodology is needed for extended sources (see Section \ref{sec:imaging}). RedCan finishes 
here if the input list of files does not include any spectroscopic data. It continues to 
the next steps otherwise.

\subsection{Wavelength calibration} \label{sec:wlcal}

RedCan uses sky emission features in the reference stacked spectra to wavelength
calibrate all the spectra (standard stars and targets). First, the sky spectrum is stacked
along the spatial axis. The pipeline looks for the local maxima in this stacked spectrum
as local peaks within the 10 closest pixels. A gaussian
profile is fitted to the 11 pixels centred in the local maxima to accurately determine the
position of the centre. This method is rather sensitive to broad features such as those
from the MIR sky.

The next step is to identify these local maxima with known features. For that purpose the
pipeline contains a set of wavelength-calibrated sky templates for the N and Q band and
for T-ReCS and CanariCam. These observational templates are obtained from the best
weather conditions data from the Gemini archive and commissioning data, for T-ReCS and
CanariCam data, respectively. After selecting the associated template according to the
grating and instrument, a grid of offsets and distortions is applied to it. The best
offset and distortion for the sky-spectrum is chosen by computing the minimum
$\chi^{2}_{r}$. Using the wavelength calibration of the templates, RedCan identifies as
many features as possible in the sky-spectrum. A re-identification of these lines along
the spatial axis is made each 10 pixels. A linear fit is then applied along the wavelength
direction and third and fourth order fit are applied along the spatial direction for T-ReCS
and CanariCam data respectively. The final calibration is stored in a 2D wavelength
calibration array.

The N-band spectral resolution is R=100 and R=175 for T-ReCS and CanariCam
respectively. The accuracy of the wavelength calibration is always much better than the
spectral resolution of the instruments. 

\subsection{Trace determination}\label{sec:trace}

Before extracting the spectra the traces of standard stars and targets are computed. The
pipeline stacks the spectrum along the wavelength axis every 10 pixels and determines the
centre and width of the trace by fitting it to a MOFFAT profile (the
mpfitpeak\footnote{http://cow.physics.wisc.edu/~craigm/idl/idl.html} function of IDL). The
resulting widths of the trace along the spatial direction are fitted to a polynomial. A
first and second order degree polynomials are tested against the N-band spectra to include
the possibility that the data might not be diffraction-limited at bluer wavelengths. For
the width and center of the trace in the Q-band a linear fit is used for both T-ReCS and
CanariCam. The centre and width of the trace for each observation are saved in its header
with the keywords {\sc trctra, trctrb, trfwhma, trfwhmb}, and {\sc trfwhmc}.

\subsection{Spectral extraction}\label{sec:extraction}

RedCan extracts the spectrum by using the 2-D final spectrum of the target and
standard star and the wavelength calibration matrix (see Section \ref{sec:wlcal}). For
the centre of the extraction region the pipeline uses the centre of the trace for each
object recorded in the header (see Section \ref{sec:trace}). For the width of the
extraction region two considerations can be made: (1) extended or (2) point-like
sources\footnote{These two options can be selected by the user.}.

For point-like sources the width along the spatial direction as a function of
wavelength is set to vary as the trace of the associated standard star. The user can
specify the number of times the width of the standard to be used. To extract the
unresolved component this number should be set to 1. For extended sources the width does
not vary with wavelength and the user can specify the number of pixels to be extracted.
Additionally, a number of offsets (in pixels) from the centre of the target (i.e. centre
of the trace) can be specified to extract one or several regions out of the centre of the
target. The pipeline can extract either several widths for an offset or several offsets
for a given width.

For our AGN sample we have selected the point-like source option for unresolved
sources using the width of the standard star to isolate the nuclear component from
the extended emission.

\subsection{Spectral flux calibration and combination of spectra}\label{sec:specfluxcal}

Each spectrum is flux-calibrated using the ratio between the observed associated standard
and its corresponding flux calibrated template given by \citet{Cohen99}. For the
extraction of the standard a radius of 4 times the sigma of the trace of the
standard is used to ensure the collection of the whole flux of the standard. 
 
For extended sources the final flux-calibrated spectrum is obtained by dividing the
spectrum (in counts) by this ratio. For point-like sources, in order to obtain a well
flux-calibrated spectrum, losses due to the width of the slit (slit-losses) and due to the
aperture used (aperture-losses) are taken into account. The slit losses are automatically
computed using the acquisition images for the associated standard star. Two images must be
present within the observations, before and after including the slit in the optical
system. The pipeline computes aperture photometry in these two images and put the ratio of
the counts of the observations without slit divided by the counts using the slit in the header
adding the new keyword {\sc slitcorr} in the observation corresponding to the spectrum of
the star. When the user selects the option point-like the final spectrum is multiplied by
this value to obtain the final flux-calibrated spectrum. Note also that slit-losses are
wavelength-dependent for point-like sources because the angular size of the PSF in a
diffraction limited observation increases linearly with wavelength. Therefore the loss
ratio calculated using the acquisition images, with and without slit, is only a first
order correction, since it is based on the averaged flux of the source as measured through
the filter used for each acquisition and thus is a constant value. While this is not
important for point-source extractions, it is a source of uncertainty for extended
sources. When observing unresolved sources, both standard and target are affected by the
same wavelength-dependent slit loss, which is automatically accounted for when the
spectrum of the target is divided by that of the standard. On the other hand, extended
sources are assumed as not having any loss due to either the slit or aperture. Therefore
the standard star introduces an uncertainty in the shape of the extracted spectrum of the
target even when the standard is corrected by the ``constant" loss ratio. 


For the aperture losses the standard star is extracted with the same width as the
target. These aperture losses depend on the wavelength and they are computed
as the ratio between the standard spectrum (with 4 times the standard trace) and the
spectrum extracted with the same trace as the target.

The errors are estimated as the sum of the statistical error (squared root of the
number of counts) plus a 15\% systematic errors due to the flux calibration procedure 
\citep[see][]{Diaz-Santos08}. 

The observations are grouped together into targets using the keyword {\sc objname}
in the headers. All the flux- and wavelength-calibrated spectra of the same target are
combined into one. This final spectrum is given at rest-frame by using the redshift
obtained from NED\footnote{http://ned.ipac.caltech.edu/}. For this the name of the target
is used. When the object is not found in NED (either because it is not extra-galactic or
the name is not well spelled in the header), a redshift equal to zero is used. Note that the
applied redshift is stored in an ascii file.

Table~\ref{tab:observations} shows the observational details for our sample. In particular
the information on the standard star used to flux-calibrate each observations and the
number of observations combined to produce the final spectrum of each target. 

\begin{figure}[!t] 

\begin{center}

\includegraphics[width=1.\columnwidth]{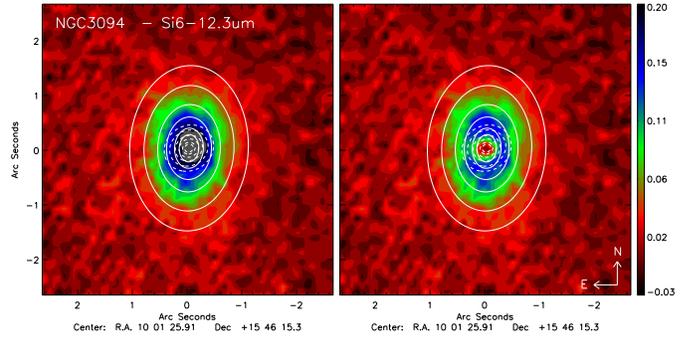}  

\caption{Si6 image at 12.3$\mu m$ of NGC\,3094 before (left panel) and after (right panel) 
the PSF subtraction. In the latter, one can clearly see the nuclear emission in red. The 
dashed contour is the best MOFFAT fit for the standard star. The color scale is in units 
of Jy.}

\label{fig:Imaging} 

\end{center} 

\end{figure}

\section{Data analysis}

\subsection{MIR images}\label{sec:imaging}

RedCan produces flux-calibrated images (see Section~\ref{sec:imagefluxcal}) using the
associated standard star images and their theoretical spectra given by \citet{Cohen99}.
Fig. \ref{fig:Imaging} (left) shows the resulting flux-calibrated image for NGC\,3094.
Most of the N-band images are acquisition images associated to the target. For that reason
most of the sources have imaging in the N-band (except Centaurus\,A, NGC\,5506, and
NGC\,5728). Q-band filter images are available for 5 objects while the narrow band filters
Si2 (8.8$\mu m$) and Si6 (12.3$\mu m$) are available for 9 and 3 sources, respectively.
All sources have an image at least in one of the filters.

Our aim is to study the central region morphology of the sample in the MIR and isolate the
AGN contribution. We have used the closest (in time) standard star observed with the same
filter as representative of the PSF of a point-like source. This star is fitted to a 2D
MOFFAT profile (using the ``mpfit2dpeak'' IDL function). We have scaled this PSF function
to the peak of the nuclear source and subtracted the nuclear component. Note that this
method would overestimate the nuclear flux of the source if extended. However, it is not
our aim to measure fluxes (they are estimated with the spectra) and this approximation is
robust at determining the number of extended sources.

The right panel in Fig.~\ref{fig:Imaging} shows the residual after the subtraction of the
PSF (dashed contours) for NGC\,3094. We consider a source as extended if the image after
the PSF subtraction shows an integrated flux higher than 3 times the standard deviation of
the background. Column 2 in Table~\ref{tab:results} indicates whether the sources are
extended (E) or point-like (P).

In general, for the sources in our sample, the central morphology is extended. Only 6 AGN
show point-like nuclear emission, namely NGC\,1365, Centaurus\,A, NGC\,5728, IC\,4518W,
ESO\,103-G35, and NGC\,7479. However, \citet{Alonso-Herrero12} found an extended
morphology for NGC\,1365 using N-band images with T-ReCS. Their proprietary data have a
higher signal-to-noise since our image is only the acquisition image for the spectrum.
Thus, the S/N influences the identification of the extended emission in NGC\,1365. In the
cases where the source is not  extended the nuclear emission comes from less than 7-236
pc, depending on the distance to the galaxies. The sources observed using several filters
show similar morphologies in all the filters. NGC\,4945 is the only AGN without an
unresolved point-like source. Overall, we find that the morphology of the nuclear source
has not clear relation with the AGN type.

\begin{figure}

\begin{center}

\includegraphics[width=1.\columnwidth]{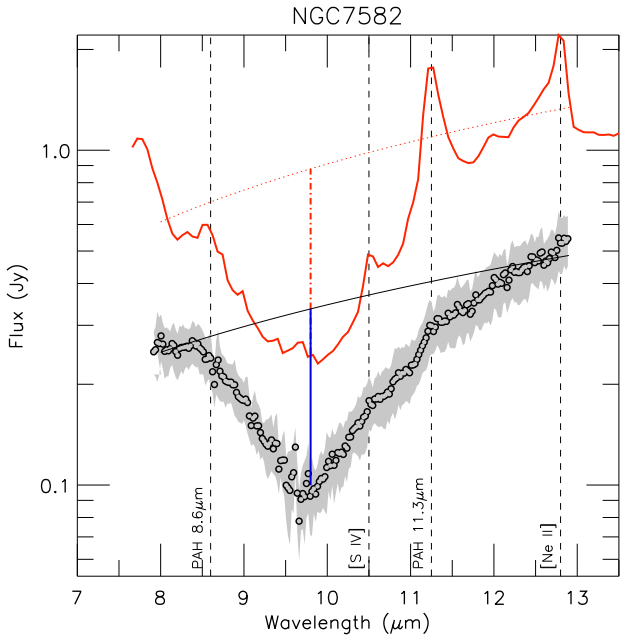}

\caption{T-ReCS (open dots) and \emph{Spitzer}/IRS (red-solid line) spectra of
NGC\,7582. The errors are shown as a grey area. Interpolated continuum and estimated 
silicate feature depth are shown as a black-solid (red-dotted) line and vertical solid 
blue (red dotted-dashed) line for the T-ReCS (\emph{Spitzer/IRS}) spectrum. The appendix A 
includes the spectra of the sample.}

\label{fig:spectrum} 

\end{center}

\end{figure}

 \begin{table*}
 \begin{center}
 \scriptsize{
 \begin{tabular}{l c c c c c c c c c c c c}
\hline\hline  	       
Name    &   E/P     & \multicolumn{3}{c}{$\tau_{9.7}$}  &  \multicolumn{2}{c}{$f_{\nu =12\mu m}(mJy)$}   & \multicolumn{2}{c}{$log(\nu L_{\nu=12\mu m}~(erg/s)~)$}  & \multicolumn{2}{c}{$EW(PAH~11.3\mu m)$}  & \multicolumn{2}{c}{$EW([SIV])$}   \\  \cline{3-5} \cline{8-9} \cline{12-13}
                    &    	    & T-ReCS       &   \multicolumn{2}{c}{\emph{Spitzer}} &  T-ReCS       &   \emph{Spitzer}   & T-ReCS       &   \emph{Spitzer}   & T-ReCS       &   \emph{Spitzer}   & T-ReCS       &   \emph{Spitzer} \\   
		    &  	  	    &              &     (linear)   &  (PAHFIT)  &      &      &      &     &    &     &          \\ \hline  
        NGC\,1365 & P  &     -0.03$\pm$      0.03 &       0.20$\pm$      0.01 &  0.00$\pm$      0.01 &     373.1$\pm$       9.8 &      752.9$\pm$       2.0 &  42.71 &  43.02 &    $<$31.3 &      113.6 &      $<$8.9 &       13.1  \\
        NGC\,1386 &  E &      0.73$\pm$      0.16 &       0.68$\pm$      0.01 &  0.53$\pm$      0.01 &  291.2$\pm$      44.3 &      498.0$\pm$       6.0 &  42.52 &  42.75 &     $<$41.8** &       47.6 &       56.3 &       32.4  \\
        NGC\,1808 & E  &      0.53$\pm$      0.03 &       1.17$\pm$      0.06 &  0.67$\pm$      0.01 &  293.7$\pm$       7.8 &      683.8$\pm$      18.7 &  42.23 &  42.59 &      183.1 &     	520.8 &     $<$27.5 &       14.6  \\
        NGC\,3081 &  E &      0.14$\pm$      0.10 &       0.24$\pm$      0.03 &  0.00$\pm$      0.01 &  177.7$\pm$      19.4 &      255.0$\pm$       2.5 &  42.91 &  43.06 &     $<$12.1 &       34.1 &       36.3 &       60.8  \\
        NGC\,3094 &  E &      3.04$\pm$      0.19 &         \dots             &  \dots					   &  734.6$\pm$      11.3 &     \dots                 &  43.80 & \dots  & $<$8.1 &      \dots &       $<$88.8 &      \dots  \\
        NGC\,3281 & E  &      1.21$\pm$      0.11 &       1.15$\pm$      0.07 &  1.29$\pm$      0.05 &  813.3$\pm$      94.1 &      691.8$\pm$      81.2 &  43.86 &  43.79 &     $<$7.7 &      \dots &        $<$5.8** &       14.6  \\
        NGC\,4418 &  E &      3.78$\pm$      0.27 &       3.59$\pm$      0.06 &  2.60$\pm$      0.05 &  735.0$\pm$      86.9 &      686.0$\pm$       2.1 &  43.49 &  43.46 &      $<$51.6 &     $<$32.2** &      \dots &       \dots  \\
        NGC\,4945 & E*  &     \dots               &       4.94$\pm$      0.08 &  4.66$\pm$      0.02 &   \dots               &      936.4$\pm$      24.0 &  \dots &  41.80 &       \dots &   286.4 &      \dots &      \dots  \\
     Centaurus\,A &  P &      1.04$\pm$      0.03 &       0.91$\pm$      0.01 &  1.10$\pm$      0.01 & 1655.0$\pm$      47.1 &     1648.6$\pm$       8.8 &  41.98 &  41.98 &      $<$9.8 &       37.2 &       20.2 &        8.1  \\
        NGC\,5135 &  E &      0.72$\pm$      0.07 &       1.01$\pm$      0.04 &  0.15$\pm$      0.01 &  126.7$\pm$       4.0 &      250.5$\pm$       1.5 &  43.27 &  43.57 &      $<$14.4 &      384.9 &       31.0 &       26.4  \\
       Circinus   & E  &      1.33$\pm$      0.09 &       1.15$\pm$      0.01 &  0.73$\pm$      0.01 & 18617.0$\pm$    1555.0 &    15579.6$\pm$     127.0 &  43.15 &  43.07 &     $<$13.1 &       39.3 &     $<$17.0 &        9.6  \\
        NGC\,5506 & E  &      0.95$\pm$      0.34 &       0.80$\pm$      0.01 &  0.71$\pm$      0.01 &  1489.0$\pm$     398.6 &     1299.6$\pm$       1.6 &  43.73 &  43.67 &     $<$13.1 &       36.6 &       13.6 &       36.3  \\
        NGC\,5643 &  E &      0.57$\pm$      0.12 &       0.67$\pm$      0.02 &  0.21$\pm$      0.01 &  277.1$\pm$      37.0 &     287.2$\pm$	 2.6 &  42.53 & \dots  &       $<$33.2** &  178.9 &       38.6 &  51.4  \\
        NGC\,5728 &  P &      \dots               &       1.22$\pm$      0.04 &  1.10$\pm$      0.01 &   \dots               &     109.6$\pm$	 1.7 &  \dots &  \dots &      $<$87.4 &     402.7 &    $<$43.7 &  197.8  \\
        IC\,4518W & P  &      1.54$\pm$      0.19 &       1.11$\pm$      0.01 &  1.10$\pm$      0.02 &  181.9$\pm$      30.0 &      283.7$\pm$	 1.0 &  43.58 & \dots  &  $<$17.8 &    104.6 &       41.2 &   64.2 \\
     ESO\,103-G35 &  P &      0.92$\pm$      0.09 &       0.83$\pm$      0.01 &  0.65$\pm$      0.01 &  578.4$\pm$      34.0 &      713.9$\pm$       5.2 &  43.91 &  44.00 &       $<$7.1 &       \dots &       23.4 &       14.8  \\
         IC\,5063 &  E &      0.39$\pm$      0.19 &       0.32$\pm$      0.02 &  0.01$\pm$      0.01 &  627.8$\pm$     127.4 &     1081.8$\pm$       2.9 &  43.81 &  44.04 &    $<$26.8 &      \dots &       22.0 &       20.8  \\
        NGC\,7130 &  E &      0.66$\pm$      0.09 &       0.87$\pm$      0.03 &  0.10$\pm$      0.01 &  148.9$\pm$      10.7 &    232.8$\pm$    4.5 &  43.49 & 43.80 &      103.8 &    225.7 &       10.9 &    15.1  \\
        NGC\,7172 & E  &      1.99$\pm$      0.35 &       1.87$\pm$      0.03 &  2.10$\pm$      0.01 &  156.2$\pm$      13.9 &      192.9$\pm$       2.1 &  42.89 &  42.98 &       $<$39.5** &      156.0 &       60.1 &       40.0  \\
          3C\,445 & E  &     -0.12$\pm$      0.04 &      -0.14$\pm$      0.02 &  0.00$\pm$      0.01 &  167.7$\pm$       7.9 &      171.4$\pm$       3.1 &  44.62 &  44.63 &      $<$22.9 &      \dots &      $<$11.6 &        4.6  \\
        NGC\,7479 &  P &      2.51$\pm$      0.18 &       2.00$\pm$      0.03 &  1.85$\pm$      0.01 &  532.7$\pm$      77.1 &      563.7$\pm$       4.6 &  43.42 &  43.44 &      $<$8.3 &       13.1** &      \dots &       15.0  \\
        NGC\,7582 & E  &      1.17$\pm$      0.21 &       1.17$\pm$      0.01 &  0.85$\pm$      0.01 &  409.1$\pm$      77.2 &     1096.0$\pm$       2.5 &  42.87 &  43.30 &       $<$23.6** &      376.7 &        $<$5.4** &       43.8  \\
\hline
\end{tabular}
\caption{Main results.$\tau_{9.7}$ is measured in the 8-13$\mu m$ band except when using
PAHFIT in the \emph{Spitzer} spectra for which we have used the full 5-35$\mu m$ band. (P)
Point-like source and (E) Extended. *NGC\,4945 has no point-like component. Equivalent
widths in units of $10^{-3}\mu m$. **Tentative detection of PAH at 11.3$\mu m$ or [SIV] at
10.5$\mu m$ (1 sigma above the continuum level).}
\label{tab:results}  
}
\end{center}
\end{table*}

\subsection{MIR spectra}\label{sec:spectra}

Fig.~\ref{fig:spectrum} shows the nuclear spectrum of NGC\,7582 (open dots and gray area).
The spectra of the entire sample are included in appendix A. The shaded regions show the
systematic errors of the extracted spectra. The spectra show systematic errors of less
than 10\% of the flux in all the objects except for NGC\,5728, in which the error is much
higher ($\sim 40$\%). We have excluded NGC\,5728 from this spectral analysis due to the
lack of good quality data. Moreover, NGC\,4945 is also excluded due to the lack of a
nuclear component. Thus, our sample of nuclear spectra comprises 20 AGN. Our sample
has three objects in common with the sample reported by \citet{Honig10a} using VISIR/VLT
data (NGC\,5643, IC\,5063, and NGC\,7582). The shape and detected features in the three
VISIR spectra in common with our sample are consistent with those reported with T-ReCS.
The total flux calibration is consistent with ours except for IC\,5063, where we find a
flux level twice fainter than that reported by Honig.

The low spatial resolution spectra (Short-Low module) \emph{Spitzer} spectra in the same
spectral range as T-ReCS are also shown in the appendix when available (red thick
line). \emph{Spitzer} data were retrieved from the Cornell Atlas of \emph{Spitzer}/IRS
Sources \citep[CASSIS][]{Lebouteiller11}. Note that these spectra are extracted using the
optimal extraction region to ensure the best signal-to-noise ratio. IC\,4518W was
reduced with the Smart package\footnote{SMART was developed by the IRS Team at Cornell
University.} \citep[version 8.2.6][]{Higdon04,Lebouteiller10} since it was not available
through CASSIS. There were \emph{Spitzer}/IRS spectra for all the AGN in our sample except for
NGC\,3094.

The observed apparent optical depth at $\lambda \sim 9.7 \mu m$ can be measured as the
ratio between the expected continuum flux ($f_{9.7,cont}$) and the observed flux in the feature
\citep[$f_{9.7,obs}$, see][]{Shi06,Spoon07,Levenson07,Georgantopoulos11}:

\begin{equation} \tau_{9.7}= ln(f_{9.7,cont}/f_{9.7,obs}) \end{equation}

Sources with absorption (emission) features show positive (negative) optical depths. To
estimate $f_{9.7,cont}$ we have fitted the continuum to a linear relation to the 8.5$\mu
m$ and 12.5$\mu m$ flux for each object. Fig.~\ref{fig:spectrum} shows the continuum  as a
solid (dotted) thin line for T-ReCS (\emph{Spitzer}) data. The $f_{9.7,cont}$ is estimated
by the linear interpolation of this continuum flux at $\rm{9.7\mu m}$ (in
Fig.~\ref{fig:spectrum} horizontal blue thick and red dot-dashed lines for T-ReCS and
\emph{Spitzer} data respectively). The error on the $\tau_{9.7}$ was determined by the S/N
of the continuum flux. More sophisticated methods as such spline fit \citep{Mason12},
PAHFIT \citep{Smith07}, or DecompIR \citep{Mullaney11} are also useful to measure
$\tau_{9.7}$ specially when the AGN is contaminated by extra-nuclear emission. However, as
discussed in the previous section our sample are mostly AGN dominated. Moreover, these
other methods require of wider wavelength ranges as they are based on a multiple component
fitting and need a larger number of data-points to avoid degeneration of the results.
However, no strong PAH features are detected in our sample (see Section
\ref{sec:contrib}). The optical depth has also been measured for the \emph{Spitzer}
spectra with the same methodology. Note that PAHFIT may result more appropriate for
\emph{Spitzer} data given its longer wavelength range. We have however used the same
method to measure $\tau_{9.7}$ as for the T-ReCS spectra for consistency in the comparison
of the silicate measurements.

\begin{figure}

\begin{center}

\includegraphics[width=1.\columnwidth]{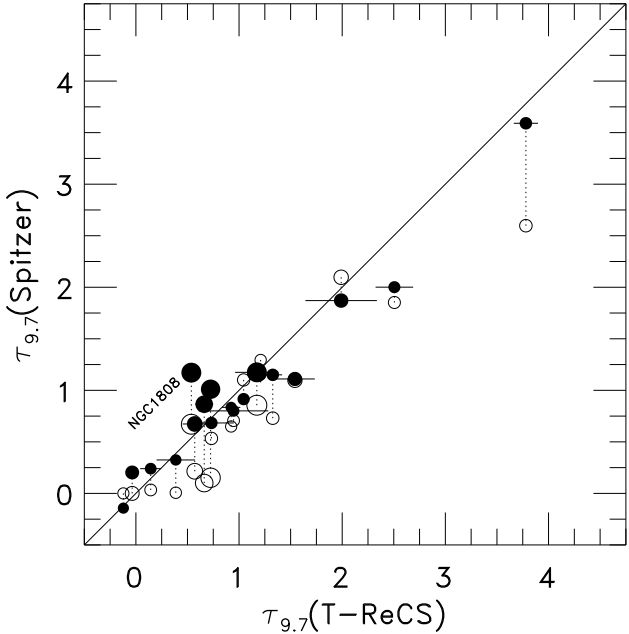}

\caption{Optical depth $\tau_{9.7}$ using \emph{Spitzer}/IRS and
\emph{T-ReCS} data. Filled symbols show the optical depth $\tau_{9.7}$
using the linear interpolation in the 8-13$\mu m$ band and empty
symbols show the optical depth $\tau_{9.7}$ using the routine PAHFIT 
in the 5-35$\mu m$ band (linked by short dotted lines). The size of the
symbols scale with the strength of the PAHs. The continuous line shows the
one-to-one relation.}

\label{fig:PAHFIT}

\end{center}

\end{figure}

Strong PAHs could introduce uncertainties in the continuum determination using our
linear interpolation \citep{Spoon07}. In order to check this, we have computed
$\tau_{9.7}$ using PAHFIT and the full wavelength range ( 5-35 $\mu m$) of the
Spitzer/IRS spectra. Fig. \ref{fig:PAHFIT} shows $\tau_{9.7}$ for \emph{Spitzer} spectra
versus that of the T-ReCS spectra. \emph{Spitzer}/IRS and T-ReCS data show similar
$\tau_{9.7}$ using the linear interpolation (see filled symbols in Fig. \ref{fig:PAHFIT}).
The mean difference between T-ReCS and Spitzer $\tau_{9.7}$ is $<\tau_{9.7}({\rm T-ReCS})
- \tau_{9.7}({\rm Spitzer})>$ = 0.03. However, when we use a broader wavelength range and
PAHFIT (see empty symbols in Fig. \ref{fig:PAHFIT}) we recover shallower optical depths
for the \emph{Spitzer}/IRS spectra ($<\tau_{9.7}({\rm T-ReCS}) -  \tau_{9.7}({\rm
Spitzer})>$ = 0.28). Thus, it seem that there is a systematic error of $\sim$
0.25 between the two methodologies.

Only in two objects the value of $\tau_{9.7}$ for the \emph{Spitzer}/IRS spectra
show a larger value than that found for T-ReCS, namely NGC\,1808 and NGC\,5135. Note that
these two sources show the largest PAH features (represented with bigger symbols in Fig.
\ref{fig:PAHFIT}). In fact, we recover a lower value for both objects when using
PAHFIT. In the following we will use the $\tau_{9.7}$ obtained by the linear
interpolation for consistency with the T-ReCS data, although in Section \ref{sec:pah} we
discuss how $\tau_{9.7}$ could affect the detection of the PAH feature.

The apparent optical depths $\tau_{9.7}$ measured are included in Table~\ref{tab:results}.
Note that in NGC\,3094 and NGC\,7479 the bottom of the silicate absorption is very noisy
so the determination of $\tau_{9.7}$ is poorly constrained. Thus, our measurement can be
considered as lower limit.


These T-ReCS data have already been presented in the literature for a small number of
individual objects. The optical depth at $9.7\mu m$ of NGC\,3281 was already reported
recently by \citet{Sales11} using the same data. They found a value of
$\rm{\tau_{9.7}=1.3\pm0.1}$, in good agreement with our results. \citet{Diaz-Santos10}
also computed the silicate absorption of three objects using T-ReCS data finding similar
results.

Most of the \emph{Spitzer}/IRS spectra give similar $\tau_{9.7}$ values {to those measured
from the T-ReCS data} although the former tend to be smaller. When using PAHFIT and
broader wavelength range these discrepancies become larger. This might reflect different
dust configurations (see Section \ref{sec:pah}). Large discrepancies are found only
for NGC\,1808, NGC\,5135, IC\,4518W and NGC\,7479. For  NGC\,1808 and NGC\,5135 the
$\tau_{9.7}$ estimates are lower than the \emph{Spitzer}/IRS results. These two sources
show the most contaminated \emph{Spitzer} spectra of the sample (strong PAH features).
NGC\,1808 shows a lower value when using PAHFIT while NGC\,5135 shows an even larger
value. 

We observed a wide range of silicate absorptions ($\tau_{9.7}$=[-0.12,3.78]) with a mean
value and a standard deviation of $<\tau_{9.7}>= 1.20 \pm 1.05$. Only two sources show
$\tau_{9.7}<0$ indicating silicate features in emission. Among them the silicate feature
seen in emission for NGC\,1365 is seen in absorption by \emph{Spitzer}/IRS. The
comparison with the literature is rather difficult due to the different criteria to
determine the continuum. However, we have compared $\tau_{9.7}$ for \emph{Spitzer} with
the $12\mu m$ sample reported by \citet{Wu09}, finding a good agreement for the 8 objects
in common.

The most significant features in the T-ReCS spectra are [SIV] at 10.5$\mu m$, the PAH
features at 8.7 and 11.3$\mu m$, and [NeII] at 12.8$\mu m$\footnote{The short wavelength
range prevent us to confirm the presence of the PAH feature at 8.7$\mu m$. Moreover, we
have not attempted to measure the [NeII] line because in most cases it is too close to the
limit of the wavelength range. }. We have measured the equivalent width (EW) of [SIV] and
PAH at 11.3$\mu m$ dividing the integrated flux of the features by the adopted continuum
near the feature (``feature'' routine within IDL). We have computed the upper
limit of the EWs accounting for the error in the continuum. We have also obtained EWs
for these lines using \emph{Spitzer} spectra. These EWs are also included in
Table~\ref{tab:results}. The continuum luminosity at 12$\mu m$, $\nu L_{\nu}(12\mu m)$,
obtained for T-ReCS and \emph{Spitzer} is also included in Table~\ref{tab:results}.

\section{Contributors to the MIR spectra of AGN}\label{sec:contrib}

\subsection{The 11.3$\mu m$ PAH feature}\label{sec:pah}

The presence of PAHs has been associated to regions of star-formation, given the global
correlations between star formation activity and PAH strength
\citep[e.g.][]{Roche91,Smith07,Wu09,Diamond-Stanic10}. This emission is thought to
originate in photo-dissociation regions (PDRs) where aromatic molecules are shielded from
the radiation field produced by hot stars.

It has been proposed that the hard radiation field produced by the AGN may destroy the PAH
molecules \citep{Genzel98,Siebenmorgen04}, as indicated by the decrease of the
EW(PAH~11.3$\mu m$) with infrared colors \citep{Wu09}. Nonetheless,
\citet{Diamond-Stanic10} found 11.3$\mu m$ emission in most of their AGN in a sample of 35
Seyferts observed with \emph{Spitzer} at scales of 1 kpc,  finding suppression of the PAH features
only for the 6.2, 7.7, and 8.6 $\mu m$ PAHs \citep[see also][]{Diamond-Stanic12}.

A strong PAH feature at 11.3$\rm{\mu m}$ is present only in NGC\,1808 and NGC\,7130 in the
T-ReCS spectra. A hint of the PAH feature is also present in other 4 AGN (NGC\,1386,
NGC\,5643, NGC\,7172, and NGC\,7582). Thus, for 14 out of 20 sources their high spatial
resolution T-ReCS spectra are free of signs of star-formation in the MIR, as
traced by the PAHs.

From visual inspection of the 2D T-ReCS spectra of these sources, the PAH feature is
spatially extended for NGC\,1808 and coincident with the nuclear emission for NGC\,7130.
Moreover, the central UV continuum of NGC\,7130 consists of several knots of
star-formation distributed in an asymmetric ring within 1 arcsec, not spatially resolved by our data
\citep{Gonzalez-Delgado98}. Thus, the PAH features detected in NGC\,7130 could be
associated to this ring. NGC\,1808 shows a circumnuclear starburst region
\citep[e.g.][]{Collison94} together with an AGN \citep[][]{Veron-Cetty85}. The 3.3 $\mu$m
PAH emission of NGC\,1808 was studied by \citet{Tacconi-Garman05} using high spatial
resolution images obtained with the ISAAC (VLT) instrument. They found PAH emission
coincident with the nucleus supporting these results.

\citet{Wu09} described a method to estimate the AGN contribution when the PAH feature at
11.3$\mu m$ is present. It is based on the constant ratio between the flux of the PAH
feature and the continuum at $12 \mu m$ ($F_{12\mu m, SF} \simeq 22.73 \times F_{PAH}$)
where the fraction of the AGN at MIR is $F_{AGN}= (F_{12\mu m} - F_{12\mu m, SF})
/F_{12\mu m}$. We have computed this value for the two objects with the strongest PAH
emissions finding 87\% and 93\% for NGC\,1808 and NGC\,7130, respectively. Therefore,
the AGN contribution seems to be high even for the objects with the strongest PAH
features. However, this method shows a 25\% of error (see Section \ref{sec:LXLMIR} for
further analysis on the AGN contribution of these two sources).

\citet{Goulding12}, using \emph{Spitzer}/IRS spectra, found that star-formation
dominates in most Compton-thick AGN. They argued that this might be an intrinsic
characteristic of heavily absorbed AGN. Our sample includes 8 Compton-thick AGN but only
NGC\,7130 shows PAH emission in the T-ReCS spectra. However, as explained above, the
T-ReCS spectrum of NGC\,7130 includes an unresolved ring of star-formation. Thus, it seems
that the nuclear star-formation among Compton-thick AGN might be as uncommon as for the
rest of the AGN. However, this needs to be confirmed for a larger sample of Compton-thick
AGN.

The lack of PAHs in the nuclear spectral spectra of AGN was already claimed by
\citet{Siebenmorgen04} using TIMMI2 spectra. \citet{Honig10a} also reported a lack of PAH
features in their high-spatially resolved VISIR spectra, detecting some minor emission of
PAH at 11.3$\mu m$ only in 5 cases (in the two objects in common we also detect this
tentative PAH emission).

When we compare the T-ReCS and the \emph{Spitzer}/IRS spectra, for those objects with
strong PAH detected with T-ReCS, (i.e. NGC\,1808 and NGC\,7130) the \emph{Spitzer}
PAH 11.3$\mu m$ feature is indeed present. Besides the PAH feature is detected in 16
other objects. The \emph{Spitzer} spectra are free of PAH emission lines in only 4 objects,
named NGC\,3281, ESO\,103-G35, IC\,5063, and 3C\,445. It seems that in most cases
the PAH emission comes from emitting regions outside of 7-130 pc (depending on the
distance) to the AGN. Considering that the \emph{Spitzer} spectra have a spatial scale of
3 arcsec, that extension corresponds to anything between 0.6--0.8 arcsec in T-ReCS (depending
on observation) and the 3 arcsec used for \emph{Spitzer} spectra. This is consistent with
previous studies that suggested that AGN lack strong PAH features
\citep[][]{Genzel98,Hernan-Caballero09,Honig10a}.

However, note that the PAH at 11.3$\rm{\mu m}$ is embedded in the silicate absorption
feature and thus in cases of high extinction the PAH feature could be attenuated
\citep[e.g.][]{Brandl06,Pereira-Santaella10a}. According to \citet{Smith07}, the optical
depth at 11.3$\rm{\mu m}$ is related to that at 9.7$\rm{\mu m}$ as $\tau_{11.3\mu m} =
0.44 \times \tau_{9.7\mu m}$. For optical depths of $\tau_{9.7\mu m}= [1,2,3]$ (common
optical depths in our sample, see Section \ref{sec:silicate}) the intensity of the line
could be reduced by a factor of $I= I_{o}\times[0.64,0.41,0.26]$. Thus, the attenuation of
the PAH emission feature is expected to be large in some cases. Since the optical
depths measured in our sample are similar for both T-ReCS and \emph{Spitzer} spectra, we
would expect that the PAHs must be attenuated in both. However, using PAHFIT for
\emph{Spitzer} spectra (see Section \ref{sec:spectra}), the $\tau_{9.7}$ optical depths
are shallower in the \emph{Spitzer} spectra than in T-ReCS. Thus, the lack of PAH features
in the latter could be the result of the extra attenuation suffered by the T-ReCS
spectra. This attenuation can be expressed as $I= I_{o}\times exp(-0.44(\tau_{9.7} +
\Delta \tau_{9.7}))$, where $\Delta \tau_{9.7}= \tau_{9.7}({\rm T-ReCS}) - \tau_{9.7}({\rm
Spitzer})$ is the difference between the attenuation measured using T-ReCS and
\emph{Spitzer} spectra. Using the values of $\tau_{9.7}$ given in Table \ref{tab:results},
in most of the cases this extra attenuation cannot explain the lack of or weakness of the
11.3$\mu m$ PAH seen in T-ReCS data. In particular, the upper limit on the EW(PAH at
11.3$\mu m$) for T-ReCS data could be explain as a result of the extra attenuation only in
two galaxies, namely NGC\,1386 and NGC\,7479. Thus, the attenuation of the spectrum might
not be responsible for the lack of PAH features in AGN.

This lack of PAH emission in the proximity of the AGN ($< 100$ pc) indicates that the AGN
emission is able to destroy the PAH molecules \citep{Wu09,Mason07}. Alternatively,
the BH growth can continue after the gas fuel is not enough to maintain the star-formation
activity \citep[see the numerical simulations give by][]{Hopkins12}. These delays have
also been detected observationally \citep[e.g.][]{Davies07}. Therefore, the lack of PAH
emission might be the result of a delay between the onset of the star formation rate and
the black hole (BH) accretion rate. 

It is not easy to disentangle which of these two possibilities is the main reason
for the lack of PAHs in AGN. If the PAHs are destroyed by the AGN we would expect that
PAHs may survive in the vicinity of low luminosity AGN. In support of this, we detected
PAH features in the nuclear region of NGC\,1808 which is the object in our sample with the
lowest X-ray luminosity. \citet{Honig10a} found that most AGN show lower continuum in the
VISIR than in the \emph{Spitzer} data when the PAH are suppressed. They argued that this
is in favor of the lack of star-formation as the main reason for the suppression of PAHs
because the continuum of the MIR spectrum is roughly constant between VISIR and
\emph{Spitzer}/IRS if the PAH destruction is an effect of grain destruction. In our sample,
although the continuum is roughly constant for some of the objects
(e.g. Centaurus~A), in some other objects the continuum suppression in the T-ReCS
data is clear (e.g. NGC\,7582).

\subsection{The [S~IV] emission line at $10.5 \mu m$}\label{sec:SIV}

Another widely used AGN diagnostic line is the [SIV] at 10.5$\mu m$ since this line is
thought to be originated in the Narrow Line Region (NLR) \citep[][and references
therein]{Dasyra11}. However, the [SIV] at 10.5$\mu m$ line arises from ions with
ionization potential of 35 eV, similar to the optical line [OIII] at $\rm{5007\AA}$
\citep[see][]{Trouille11}. Thus, this line can be produced in star-forming regions as
well as in AGN \citep[][]{Pereira-Santaella10b}.

The [SIV] line is detected in 11 AGN and a hint of this line is also present in two other
sources in the T-ReCS data (NGC\,3281, and NGC\,7582). In all these cases it is also
detected with \emph{Spitzer}. Interestingly, \emph{Spitzer} detected this line in other
five additional sources (NGC\,1365, NGC\,1808, Circinus, 3C\,445, and
NGC\,7479)\footnote{ Note that this line is also detected in NGC\,5728 but the T-ReCS
spectra show a very low S/N (see Section \ref{sec:spectra}).}. The lack of this line
in these later objects could be due to the low S/N ratio of the T-ReCS spectra. In fact,
they are the faintest emission lines, according to their EW([SIV]) in the \emph{Spitzer}
spectra. Besides, only NGC\,1365 shows a limit inconsistent with the EW reported with
\emph{Spitzer} data. Thus we claim that the origin of the [SIV] line is consistent with being nuclear in
our sample \citep[see also][]{Pereira-Santaella11}.

\begin{figure} 

\begin{center} 

\includegraphics[width=1.\columnwidth]{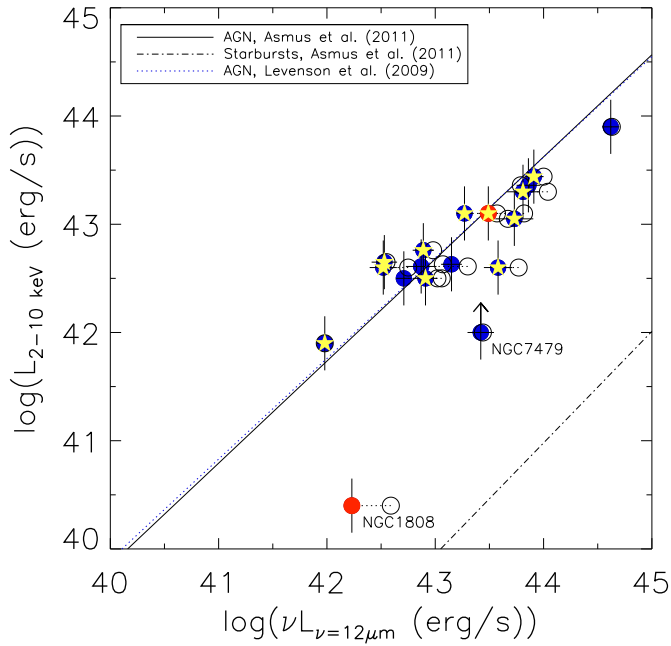}

\caption{Absorption corrected 2-10 keV luminosity versus the 12$\mu m$ luminosity, both in
logarithmic units. Filled circles are T-ReCS data and open circles are \emph{Spitzer}
data. The dotted lines link \emph{Spitzer} to T-ReCS data for the same object. Yellow
stars and red circles mark the objects with the strongest [SIV]$10.5\mu m$ and PAH at
$11.3\mu m$ emissions in the T-ReCS spectra, respectively. The blue dotted line shows the
best fit found by \citet{Levenson09} for AGN. Black dotted-dashed and solid lines
represent the best fit found by \citet{Asmus11} for Starburst galaxies and AGN,
respectively.} 

\label{fig:LxLMIR} 

\end{center} 

\end{figure}

\subsection{X-ray to MIR relation}\label{sec:LXLMIR}

The mid-IR emission in AGN is originated in the reprocessing of the soft X-ray and UV
radiation by dust located in the torus. At the same time, escaping high-energy photons
could destroy PAH molecules at sub-kpc scales \citep{Voit92}. This is supported by
the lack of PAH emission in the T-ReCS spectra (see Section \ref{sec:pah}). The
reprocessed dust emission should then produce a significant correlation between X-ray
luminosity (after correction from X-ray absorption) and MIR luminosity. This correlation
has already being found by other authors
\citep{Horst08,Levenson09,Gandhi09,Honig10a,Ichikawa12}.

Fig.~\ref{fig:LxLMIR} shows the absorption corrected 2-10 keV luminosity $L_{2-10~keV}$
versus the 12$\mu m$ continuum luminosity $L_{12\mu m}$. X-ray errors represent the
expected X-ray variability among AGN of $0.5$ dex while the MIR errors are inferred from
the photometric errors including 10\% error for the total flux calibration. \emph{Spitzer}
luminosities (empty symbols) agree very well with the \emph{T-ReCS} luminosities (filled
symbols) for most of the sources ($\rm{log(L_{12,T-ReCS}/L_{12,Spitzer})<0.4}$).

Most of the objects follow the same correlation found for AGN by \citet{Levenson09} and
\citet{Asmus11} (blue-dotted and black-solid lines, respectively). Moreover, those objects
with significantly higher MIR emission compared to X-rays are also far from the correlation
found by \citet{Asmus11} for starburst galaxies (dot-dashed line in
Fig.~\ref{fig:LxLMIR}). Note that the X-ray luminosity of NGC\,7479 is not corrected for
absorption although the object is Compton-thick. Thus, its X-ray luminosity could be up to
60 times higher than that reported here \citep[see][]{Panessa06}, in better agreement with
the correlation for AGN. Interestingly, NGC\,1808 (the closest to the starburst correlation) shows
the strongest PAH emission in our T-ReCS sample, indicating that the AGN component of this
source is not dominating the MIR emission. The locus of this object in the correlation
indicates that the fraction of star-formation must be higher than the estimated  (87\%,
see Section~\ref{sec:pah}. In fact, the existence of an AGN in the nucleus of
NGC\,1808 has been under debate \citep[e.g.][]{Phillips93,Awaki96}.
\citet{Jimenez-Bailon05} studied the X-ray emission of NGC\,1808 with \emph{XMM}-Newton
and \emph{Chandra} showing the co-existence of thermal diffuse plasma and non-nuclear
unresolved point-like sources associated with the starburst activity, together with a low
luminosity AGN. However, it is clear from our correlation that the AGN in NGC\,1808 is not
dominating the MIR emission.

Thus, despite the heterogeneous sample and the large number of highly obscured objects,
the 12$\mu m$ luminosity and the X-ray luminosities are still well correlated. Thus,
except for the nucleus in NGC\,1808, the AGN dominates the continuum emission at MIR
frequencies.

\section{The silicate absorption/emission feature}\label{sec:silicate}

\citet{Wu09} based in \emph{Spitzer}/data found that the silicate feature in Type-1
Seyferts is rather weak while in Type-2 Seyferts it is more likely to display strong
silicate absorption.

In the T-ReCS spectra, only the spectra of NGC\,1365 and 3C\,445 show silicate
features in emission while NGC\,3081 shows very little absorption (see appendix A and
$\tau_{9.7}$ in Table \ref{tab:results}). The other sources (i.e. 17 out of 20) show very
strong silicate absorptions. Note that NGC\,1365 and 3C\,445, the two objects showing
the silicate feature in emission, are Type 1.5 AGN. All the Compton-thick AGN in our
sample ubiquitously show strong silicate absorption feature. This is in contrast with
\citet{Goulding12} who found that most of the Compton-thick AGN in their sample do not
show such strong silicate absorption. However, since our sample it is not complete, the
statistics may well be biased by the selection of the individual observing proposals of
the targets in our sample. In particular none of the Compton-thick AGN with low optical
depth reported by \citet{Goulding12} have been observed by T-ReCS while we do share
three Compton-thick sources with strong silicate absorption features\footnote{ Note
that we have five sources in common with \citet{Goulding12} although two of them did not
have high S/N T-ReCS data for the spectral analysis.}.

\begin{figure}[!t] 

\begin{center} 

\includegraphics[width=1.0\columnwidth]{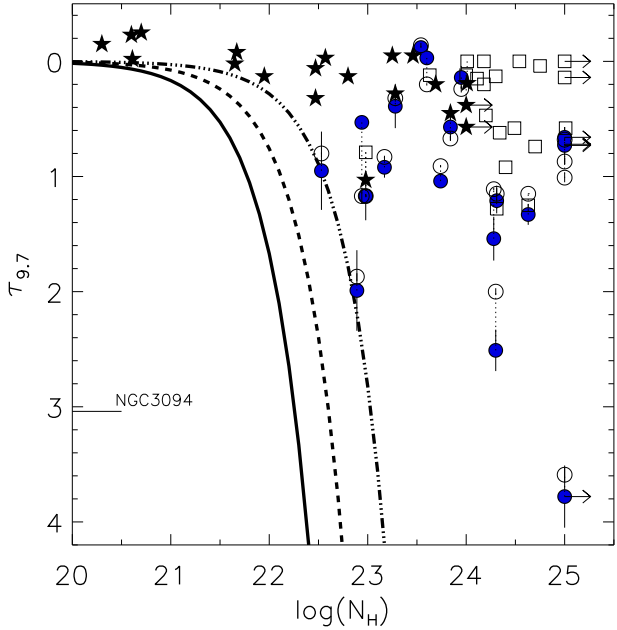}

\caption{The optical depth at 9.7$\mu m$, $\tau_{9.7}$ versus the logarithmic of the
hydrogen column density at X-rays, $\rm{N_{H}}$. $\rm{N_{H}}$ in units of $cm^{-2}$.
T-ReCS data are shown as blue-filled circles and the \emph{Spitzer}/IRS data included in
our work are shown as empty circles. The object without reported X-ray measurement is
labeled as a short line in the Y-axis. Low limit on the $\rm{N_H}$ are shown as right
arrows. Compton-thick sources reported by \citet{Goulding12} are shown as open squares and
AGN observed with VISIR/VLT by \citet{Honig10a} are shown as black-filled stars. Constant
gas-to-dust ratio (i.e. the uniform dusty torus) assuming the viewing angles of
$i=0,40,80^{\circ}$ are shown with the continuous, dashed, and dot-dashed lines,
respectively.} 

\label{fig:TAUNH1} 

\end{center} 

\end{figure}

The X-ray hydrogen column density ($\rm{N_{H}}$) of 20 of the sources reported in the
literature are shown in Table~\ref{tab:sample}\footnote{Note that there are not X-ray data
published for NGC\,3094.}. Fig.~\ref{fig:TAUNH1} shows $\tau_{9.7}$ versus $\rm{N_{H}}$ in
logarithmic scale. For comparison purposes we have included the high angular resolution
data for AGN observed with VISIR/VLT \citep{Honig10a} as solid stars as well as the
Compton-thick AGN (empty squares) observed with \emph{Spitzer}/IRS from the sample of
\citet{Goulding12}. Note that filled symbols are chosen to emphasize high spatial
resolution data (i.e. T-ReCS and VISIR).

If we focus on the high spatial resolution data there is a trend in the sense of
increasing $\tau_{9.7}$ for higher $\rm{N_{H}}$. This relation has already been seen by
several authors \citep[e.g.][]{Shi06,Wu09,Honig10a,Alonso-Herrero11}. Moreover, for
$\rm{log(N_H) > 23}$ the silicate absorption shows a wide range of values from low or
no absorption to deep silicate absorptions in agreement with the results presented by
\citet{Wu09}. In the next sections we will try to model this relationship under two
assumptions: (1) the uniform dusty torus model and (2) the clumpy torus models.

\subsection{Uniform dusty torus}

On the basis of a uniform (smooth), heavily-obscuring, sub parsec scale torus
\citep{Pier92,Pier93,Granato94,Granato97}, the gas/dust-rich torus is responsible for both
the MIR and the X-ray gas absorption through the gas-to-dust relation. The constant
gas-to-dust ratio can be estimated assuming $N_{H}/A_{V}=1.93\times 10^{21}cm^{-2}$ and
$\tau_{9.7}=0.07\times\tau_{X}$ \citep[][]{Draine07}, where $\tau_{X}$ is the total
optical depth at X-rays. The solid line in Fig.~\ref{fig:TAUNH1} shows this constant
value. For very small values of $\rm{N_{H}}$ it predicts no absorption, close to the
actual values. Note however that some objects do show the silicate feature in emission,
which cannot be explained under these models. Nonetheless, as already shown by
\citet{Goulding12}, this simple model is far from the measurements for most of the objects.

While the MIR emission is integrated over the whole dusty torus, the X-ray absorption is
due only to the amount of absorbing column density along the LOS. The influence of the
$\rm{N_{H}}$ within the LOS can be estimated assuming a density profile at the viewing
angle $i$ (angle between the LOS and the plane of the torus). \citet{Nenkova08a} assumed a
Gaussian distribution $e^{-\frac{i^{2}}{\sigma^{2}}}$, where $\sigma$ is the width of the
toroidal distribution. The constant gas-to-dust ratio for $i = [0,40,80]^{\circ}$ degrees
is shown for $\sigma=45^{\circ}$ (solid, dashed, and dotted-dashed black lines in
Fig.~\ref{fig:TAUNH1}, respectively). Only two of the sources could be explained by this
relation (NGC\,7172 and NGC\,5506), assuming a highly inclined torus. Thus the observed
$\tau_{9.7}$ feature is much shallower than predicted by this constant gas-to-dust relation.
This was already mentioned by \citet{Georgantopoulos11}, showing that although $\sim$70\%
of the sources with high optical absorption at 9.7$\mu m$ are classified as Compton-thick
AGN, the opposite is not true, since many Compton-thick sources do not show significant
absorption at $9.7\mu m$ (see Fig.\ref{fig:TAUNH1}).

\subsection{Clumpy torus} \label{sec:CLUMPY}

The clumpy torus models of \citet{Nenkova08a,Nenkova08b} predict
the silicate feature in emission even for high values of the optical depths of the individual clouds
($\rm{\tau_{V}>100}$). This effect can also explain the low silicate absorption observed for
some Compton-thick sources in our sample. In order to test this possibility we have used
the libraries of CLUMPY\footnote{http://www.pa.uky.edu/clumpy} models, a code modeling the AGN
clumpy dust torus emission \citep{Nenkova08a,Nenkova08b}. We have obtained a set of
spectral energy distributions (SEDs) for $\sigma=45^{\circ}$, $Y=R_{out}/R_{inner}= 200$
(where $R_{inner}$ and $R_{out}$ are the inner and outer radii of the torus,
respectively), an exponential slope of the radial distribution of clouds $q=2$
\citep[see][for more details]{Nenkova08b}, $\tau_{V}=5-150$, and the number of clouds
along the equator of the torus $N_{o}= 2-25$ clouds. We chose $Y$ to use the lowest maximum
temperature and therefore the maximum silicate depth. Hereinafter we assume $\sigma =
45^{\circ}$. \citet{Alonso-Herrero11} has measured $\sigma = [30-60]^{\circ}$ for a sample
of Type-2 Seyferts using a Bayesian method to fit the CLUMPY models \citep[see
also][]{Asensio-Ramos09}. Note that this parameter is not crucial since the silicate
absorption is insensitive to the angular size of the torus when the LOS incidence angle of the
ray is smaller or close to the size of the torus \citep[i.e. Type-2 Seyferts,
see][]{Nenkova08b}. Under this approximation, the total optical depth produced by the 
torus is the sum of the optical depths of the individual
clouds along the LOS ($\tau_{X}= N\times\tau_{V}$, where $N$ is the number of clouds along
the LOS) and the number of clouds $N$, similarly to the classical torus, is

\begin{equation} N=N_{o}e^{-\frac{i^{2}}{\sigma^{2}}} \end{equation}

\begin{figure}[!t] 

\begin{center} 

\includegraphics[width=1.0\columnwidth]{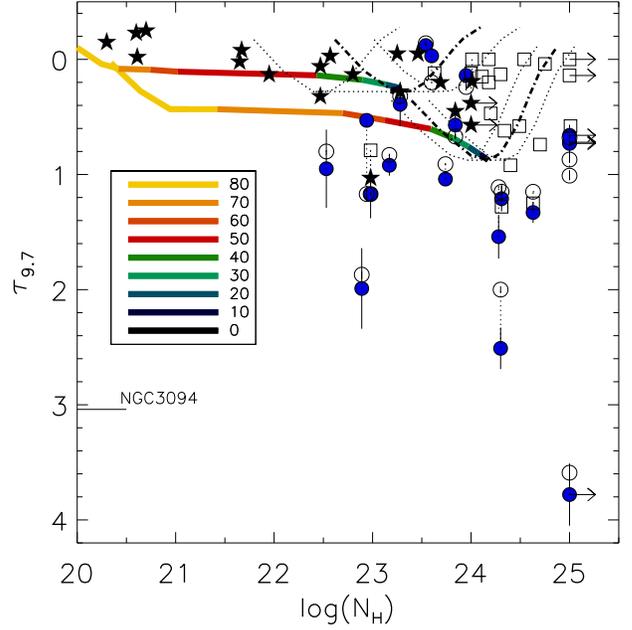}

\caption{The optical depth at 9.7$\mu m$, $\tau_{9.7}$ versus the logarithmic of the hydrogen
column density at X-rays, $\rm{N_{H}}$. $\rm{N_{H}}$ in units of $cm^{-2}$. Symbols as in
Fig. \ref{fig:TAUNH1}. Dot-dashed lines are the CLUMPY model predictions for $i =
0^{\circ}$ and $N_{o}=[5,25]$ clouds. Dotted lines show how the curves evolve for the
minimum and maximum number of clouds with a probability higher than 10\% of intercepting
the LOS. The colored thick line show how the curve evolve when the viewing angle changes
from $i=0^{\circ}$ (black) until $i =80^\circ$ (yellow). The dotted lines are the model
including the Poisson probability to detect a number of clouds along the ray (see text).} 

\label{fig:TAUNH2} 

\end{center} 

\end{figure}

We have applied the same technique described in Section~\ref{sec:spectra} to obtain the
silicate feature strength for these SEDs. Fig.~\ref{fig:TAUNH2} (dot-dashed black lines)
shows $\tau_{9.7}$ for these models for an inclination $i=0$ (i.e. along the plane of the
torus) and $N_{o}$ of 5 and 25 clouds.

Nonetheless, this number of clouds $N$ is still an estimate of the mean number of clouds
along the LOS. Statistically for a viewing angle $i$ and a mean equatorial number of
clouds $N_{o}$ the ray might intercept a range of clouds that might differ substantially
from the average. We have estimated the range of $N$ following the prescription given by
\citet{Nenkova08a} (see their Appendix). Using Poissonian distributions the probability
that a ray intercepts a number of clouds $k$ normalized to the probability to intercept
the mean value of clouds $N$ is:

\begin{equation} \frac{P_{k}}{P_{N}} = \frac{N!}{k!}N^{(k-N)} \end{equation}

\noindent the range of the number of clouds ($k$) is obtained as the minimum and maximum
for which the probability is smaller than 10\%. The dotted lines toward the higher
(lower) $N_H$ in Fig.~\ref{fig:TAUNH2} represents the expectation for the largest (lowest)
number of clouds with a probability above 10\%.

Many of the Compton-thick AGN are well within the boundaries of this model if we are
observing them close to the plane of the torus and the number of clouds is high
($N_{o}\sim5-25$). The colored thick lines show how this model changes with the
inclination angles for the two $N_{o}$. The model reproduces well all the mildly
obscured X-ray objects (with $N_{H} <10^{22} cm^{-2}$) assuming a mean number of clouds of
$N_{o}>5$. This is in good agreement with the parameters of the Bayesian fit to CLUMPY models
reported by \citet{Ramos-Almeida09}, \citet{Ramos-Almeida11}, and
\citet{Alonso-Herrero11}.

However, the model fails to reproduce a dozen sources with high value for
$\tau_{9.7}$. \citet{Nenkova08a} showed that irrespective of the optical depth, the
absorption feature produced by a clumpy torus is never deep ($\tau_{9.7}<1$). This is
confirmed by other clumpy models developed \citep[e.g.][]{Honig10b}.
\citet{Alonso-Herrero11} also noted that it is possible that for the Seyfert galaxies with
the deepest silicate features a clumpy medium in a torus-like configuration may not be
appropriate to explain the observations. \citet{Levenson07} showed that deep silicate
features require that the source is embedded in dust that is both optically \emph{and}
geometrically thick. They showed that a geometrically thick shell structure is able to
produce $\tau_{9.7}$ as high as 4. One possibility then, is that these objects show 
dust screen geometries in addition to the clouds of the clumpy torus.

\subsection{Contribution from extended dust to the observed silicate feature}

\begin{figure} 

\begin{center} 

\includegraphics[width=1.0\columnwidth]{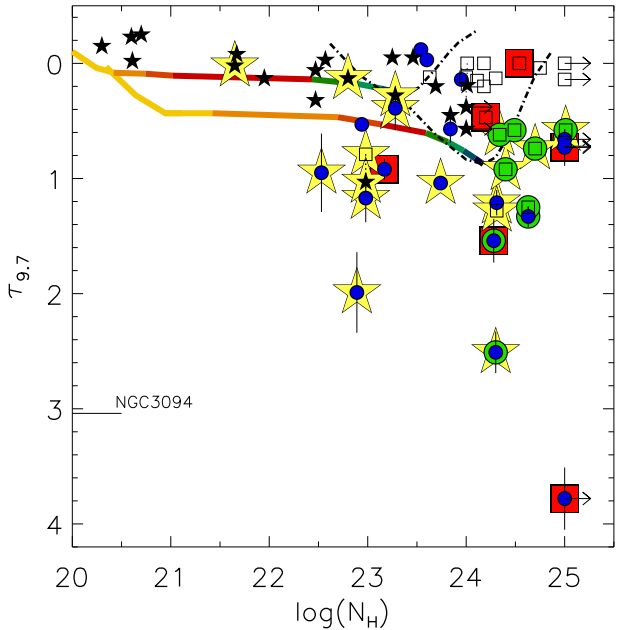}

\caption{The optical depth at 9.7$\mu m$, $\tau_{9.7}$ versus the logarithmic of the hydrogen
column density at X-rays, $\rm{N_{H}}$. $\rm{N_{H}}$ in units of $cm^{-2}$.
\emph{Spitzer}/IRS data of our sample are removed here for clarity. Edge-on galaxies are
shown as big red squares, galaxies with dust-lanes are shown with big yellow stars, and
mergers are shown as big green circles. Note that edge-on galaxies are shown only when
dust-lanes are not present for clarity of the plot.} 

\label{fig:TAUNH3} 

\end{center}

\end{figure}

On the basis of the uniform dust torus models produced from radiative transfer theory
\citep[][]{Dullemond05,Schartmann05,Fritz06}, the MIR spectrum of obscured AGN included in
our sample are expected to be dominated by a power-law and significant silicate absorption
features at $\rm{\lambda \sim 9.7 \mu m}$. However, as discussed in the previous sections,
many AGN in our sample show silicate absorption features shallower than predicted for
the homogeneous dusty model and often deeper than predicted for the clumpy model.
Several authors have claimed that the dust seen at MIR frequencies might not be co-spatial
with the torus gas measured at X-rays \citep[][]{Sturm05,Goulding12}. In particular, the
inclination angle of the host galaxy, dust lanes, or galaxy minor or major mergers may
have significant effects on the AGN emission at MIR
\citep{Malkan98,Matt00,Deo07,Deo09,Alonso-Herrero11,Goulding09,Goulding12}. Although not
many indications of extended contributors to the emission can be seen in our spectra, most
of the sources in our sample are extended  (see Section \ref{sec:imaging}). In this
section we discuss the possible contribution of the host galaxy to the optical depth at
9.7$\mu m$.

To study a possible extra-nuclear origin for the silicate absorption feature we have
looked for the superb spatial resolution that \emph{HST} can achieve. Broad band
\emph{HST} images are available in the MAST archive\footnote{http://archive.stsci.edu/}
for 12 AGN in our sample (10 with the F606W filter and 2 with the F814W filter).
Narrow-band filters have been used in 5 cases whenever broad-band filter images were not
found. For the remaining 5 objects without \emph{HST} images, we have used the available
catalog of images in NED\footnote{http://ned.ipac.caltech.edu} (NGC\,3094, NGC\,4418,
Circinus, IC\,4518W, and 3C\,445). We have compiled the axial ratio of the galaxies as the
\emph{b/a} ratio, where \emph{b} and \emph{a} are the minor and mayor axis respectively,
to measure the inclination of the host galaxy (see Table \ref{tab:sample}). Axial ratios
of $b/a < 0.5$ are considered as edge-on galaxies. Dust lanes or extended dust across the
nucleus are present in 8 objects and the AGN is hosted in edge-on galaxies in 9 objects.
In 10 objects no particular inclination or dust features are found. We have also checked
for the morphological properties of the host galaxies of the sample. Five galaxies are in
mergers, namely NGC\,4945 \citep{Aalto91}, Circinus \citep{Aalto91}, IC\,4518W
\citep{Vorontsov-Velyaminov77}, NGC\,7130 \citep{Aalto91}, and NGC\,7479
\citep{Laine99,Martin00}. These properties are included in Table \ref{tab:sample}. Note
that for comparison purposes the same properties hava been looked for in the
\citet{Goulding12} and \citet{Honig10a} samples.

Fig.~\ref{fig:TAUNH3} shows $\tau_{9.7}$ versus $\rm{N_{H}}$ where we mark those AGN
hosted in edge-on galaxies as red squares, AGN with nuclear dust-lanes as yellow stars,
and mergers as green circles. For clarity we show only edge-on galaxies without nuclear
dust-lanes. We note that all the objects with high $\tau_{9.7}$ are hosted in galaxies meeting one or
more of these three properties. In general, clumpy torus models can only
explain face-on galaxies free of either dust lanes and/or participating in mergers,
although this analysis cannot rule out the host galaxy as the responsible for the observed
$\tau_{9.7}$, even for low $\tau_{9.7}$. Only three objects show one of these properties
without a strong ($\rm{\tau_{9.7}<0.4}$) silicate absorption features (IC\,4329A,
NGC\,3227, and NGC\,424). Therefore, \emph{we conclude that even at the high angular
resolutions (100 pc) of our observations, dust in the host galaxy can play a role on the
overall dust extinction seen at MIR frequencies, and particularly it does when deep
silicate absorption features are observed.}

\section{Conclusions}\label{sec:conclus}

Here we present an analysis of 20 high spatial resolution MIR AGN spectra observed with
T-ReCS (Gemini observatory) at scales of $\sim$100 pc. The T-ReCS data have been processed
with the pipeline RedCan. RedCan has been developed by our group for the data reduction
of imaging and spectroscopic data from the CanariCam instrument, recently commissioned at
the GTC (La Palma, Spain). This data set has been complemented with \emph{Spitzer} spectra
probing kpc scales as well as X-ray information for the sources. The main results of the
study are:

\begin{enumerate}

\item On the AGN contribution to the MIR spectrum at scales of $\sim$100 pc. We find that the
AGN dominates the MIR emission as seen in the high-spatial resolution T-ReCS spectra
presented here. This is in agreement with the relatively faint 11.3$\mu m$ PAH emission
observed in most of the AGN studied here.

\item On the origin of the silicate feature. The observed strength of the nuclear (a few
hundred pc) $9.7\mu m$ optical depth could be well fitted with clumpy models with a
$\rm{N_{o}\sim 5-25}$ number of clouds but only for sources with little of mild
silicate absorption. However, sources with the deepest $9.7\mu m$ optical depths cannot be
reproduced with clumpy models. In these cases we have found that the host galaxies
contribute to the nuclear component since they are detected in objects showing optical
dust lanes and/or they are involved in a merger process.

\item On the importance of the spatial resolution. Intense 11.3$\mu m$ PAH feature emission is
detected in most of the \emph{Spitzer}/IRS spectra. This indicates that, at least
in our sample of AGN, most of the star formation activity, as probed by this tracer is
taking place in regions at least 100 pc away from the AGN. Interestingly, in most of
the AGN the 12$\mu m$ \emph{Spitzer} continuum luminosity match those observed by
T-ReCS. 

\end{enumerate}


\begin{acknowledgements} 
The authors acknowledge the Spanish MINECO through project Consolider-Ingenio 2010 Program
grant CSD2006-00070: First Science with the GTC
(http://www.iac.es/consolider-ingenio-gtc/). This work was also partially funded by the
Spanish MINECO through a Juan de la Cierva Fellowship. Based on observations obtained at
the Gemini Observatory, which is operated by the Association of Universities for Research
in Astronomy, Inc., under a cooperative agreement with the NSF on behalf of the Gemini
partnership: the National Science Foundation (United States), the Science and Technology
Facilities Council (United Kingdom), the National Research Council (Canada), CONICYT
(Chile), the Australian Research Council (Australia), Ministerio da Ciencia,
Tecnolog\'{i}a e Inovacao (Brazil) and Ministerio de Ciencia, Tecnolog\'{i}a e
Innovaci\'{o}n Productiva (Argentina). The Cornell Atlas of \emph{Spitzer}/IRS Sources
(CASSIS) is a product of the Infrared Science Center at Cornell University, supported by
NASA and JPL. C. M. Telesco acknowledges support from NSF grant AST-0903672. A.A.-H. and
P.E. acknowledge support from the Spanish Plan Nacional de Astronom\'{i}a y
Astrof\'{i}sica under grant AYA2009-05705-E. A.A.-H. is also supported by the Universidad
de Cantabria through the Augusto Gonz\'alez Linares Program and by AYA2010-21161-C02-01.
C.R.A. acknowledges the Estallidos group through project PN AYA2010-21887-C04.04. 

\end{acknowledgements}


\newpage
\onecolumn

\appendix

\section{Catalogue of spectra}

\begin{figure}[!b]
\begin{center}
\includegraphics[width=0.3\columnwidth]{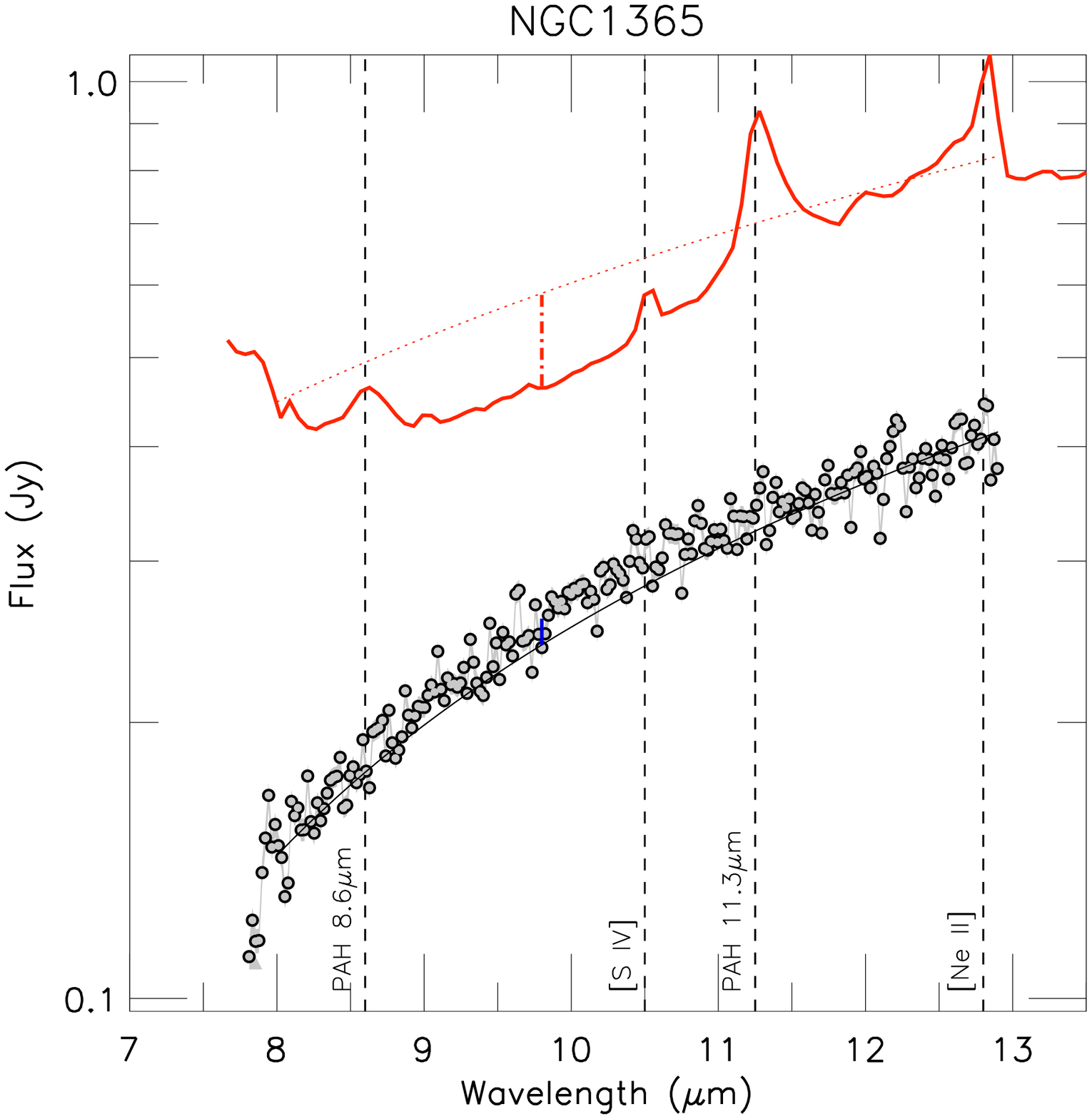}
\includegraphics[width=0.3\columnwidth]{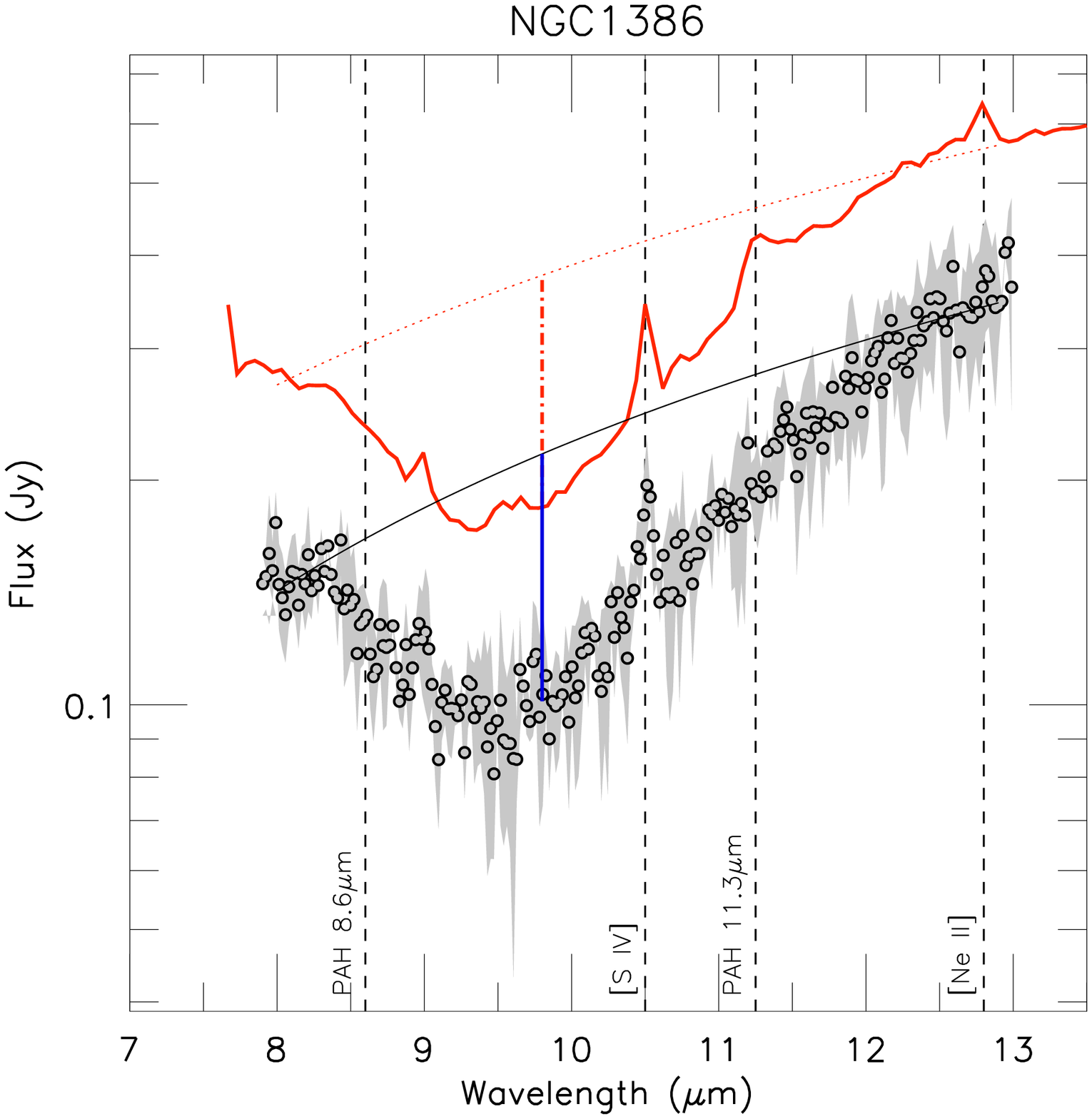}
\includegraphics[width=0.3\columnwidth]{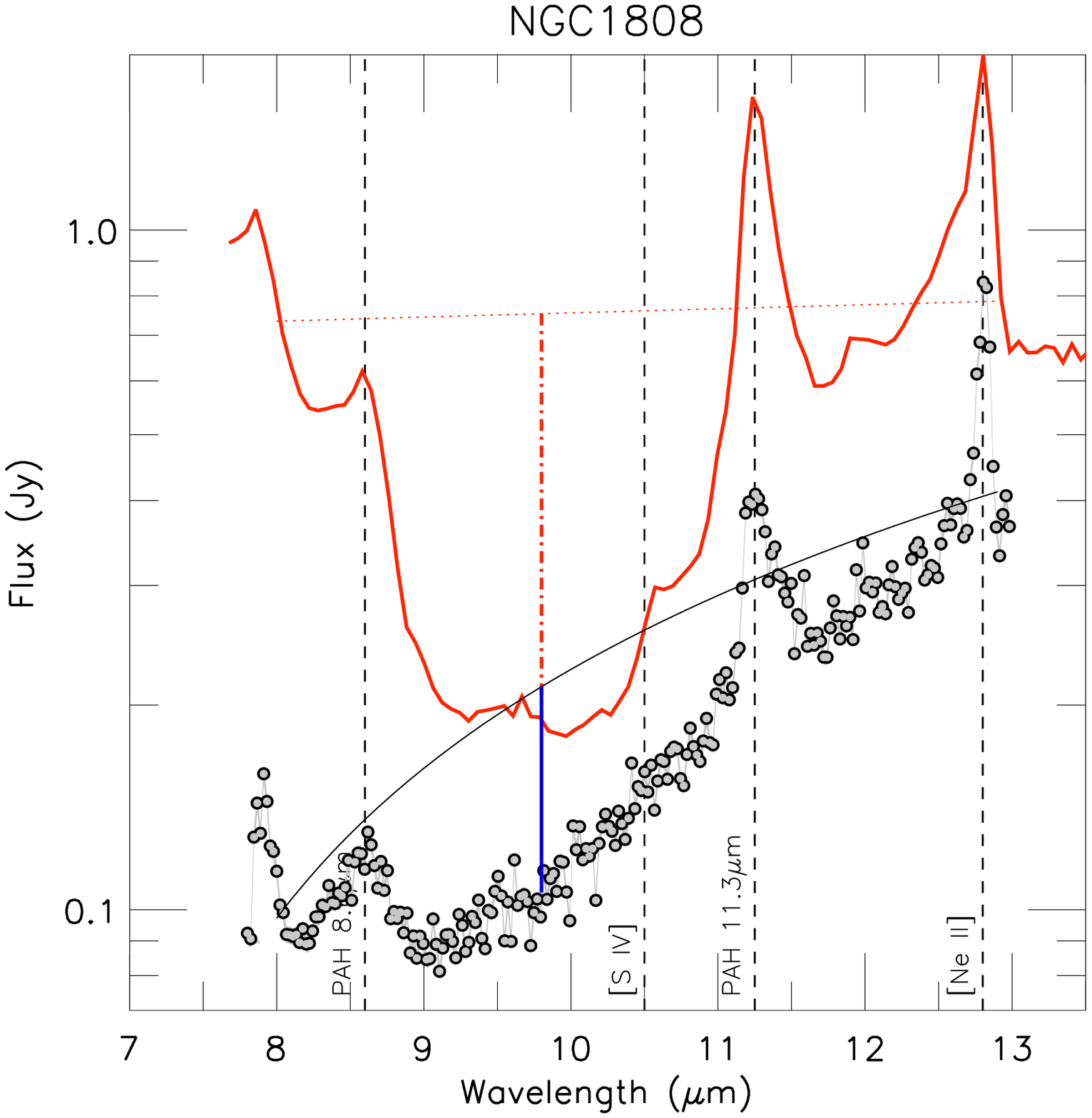}
\includegraphics[width=0.3\columnwidth]{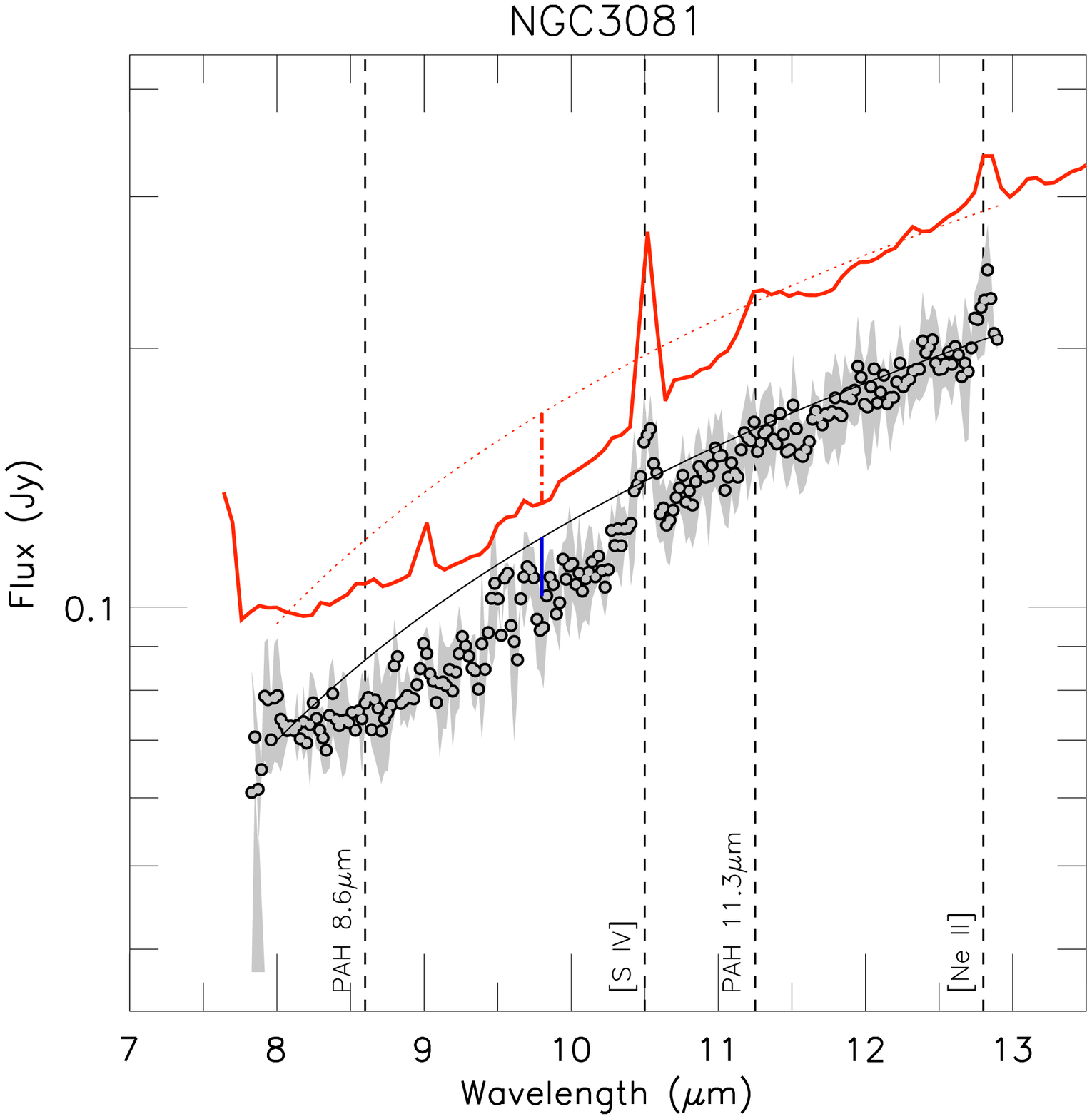}
\includegraphics[width=0.3\columnwidth]{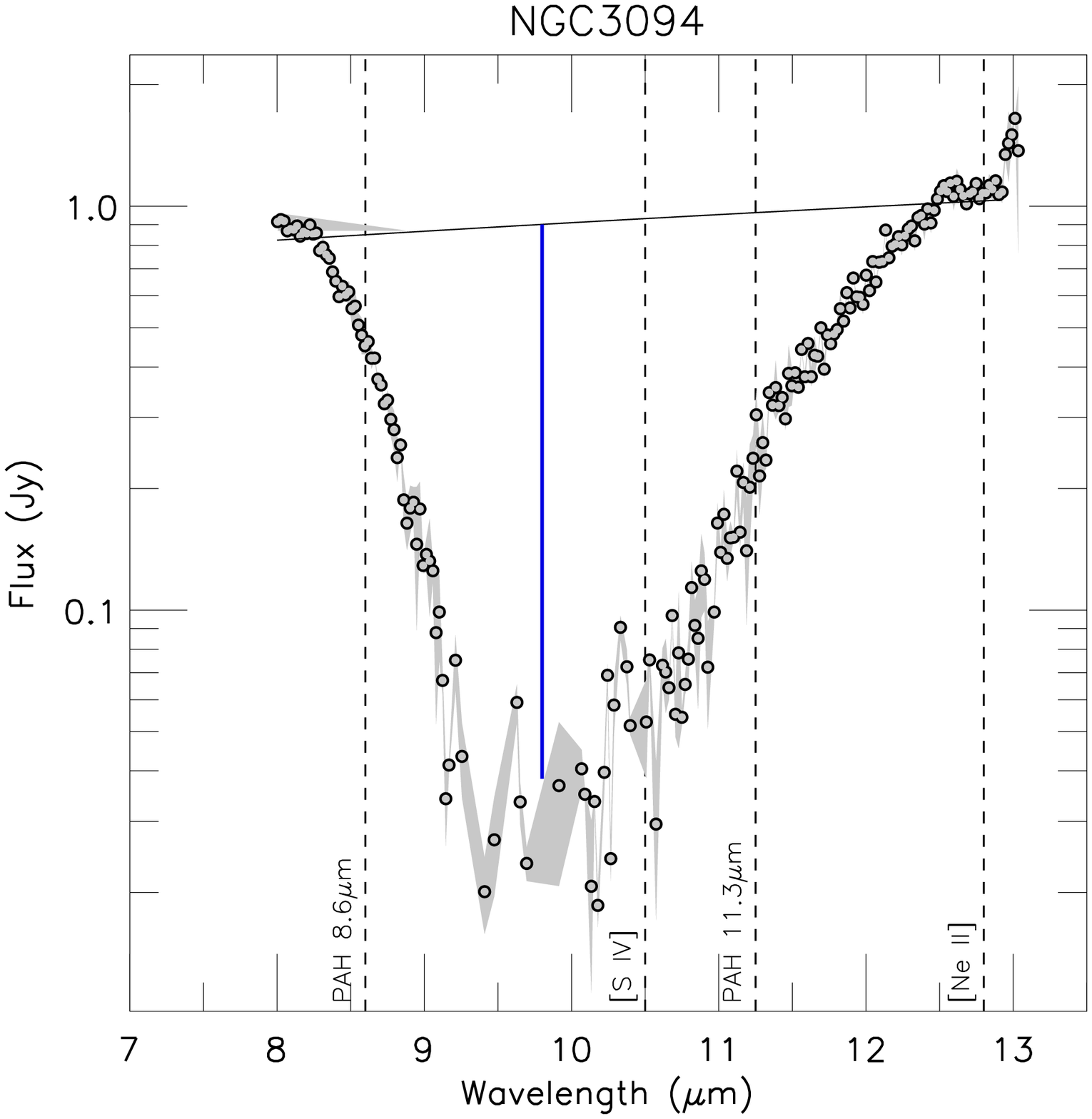}
\includegraphics[width=0.3\columnwidth]{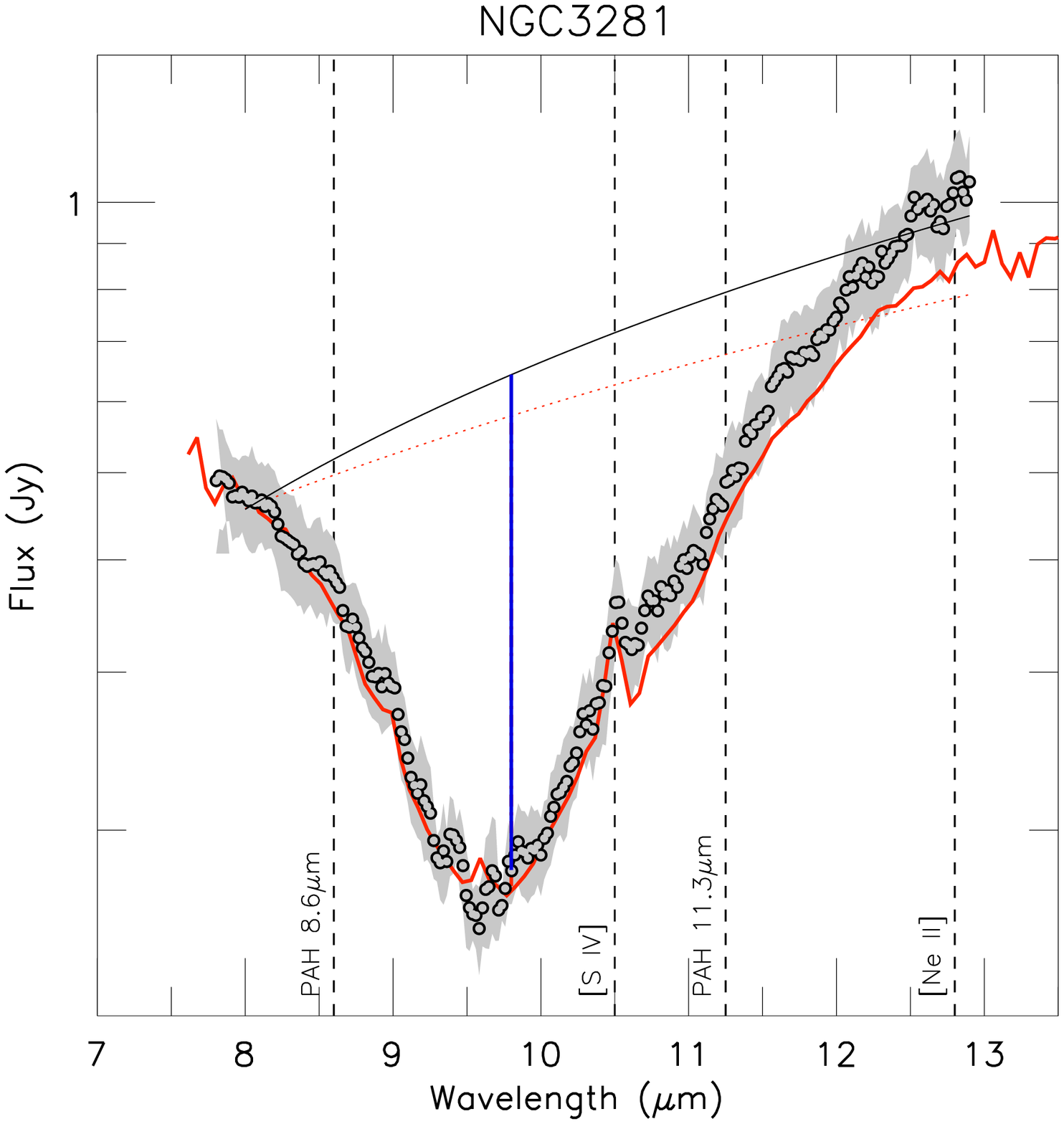}
\includegraphics[width=0.3\columnwidth]{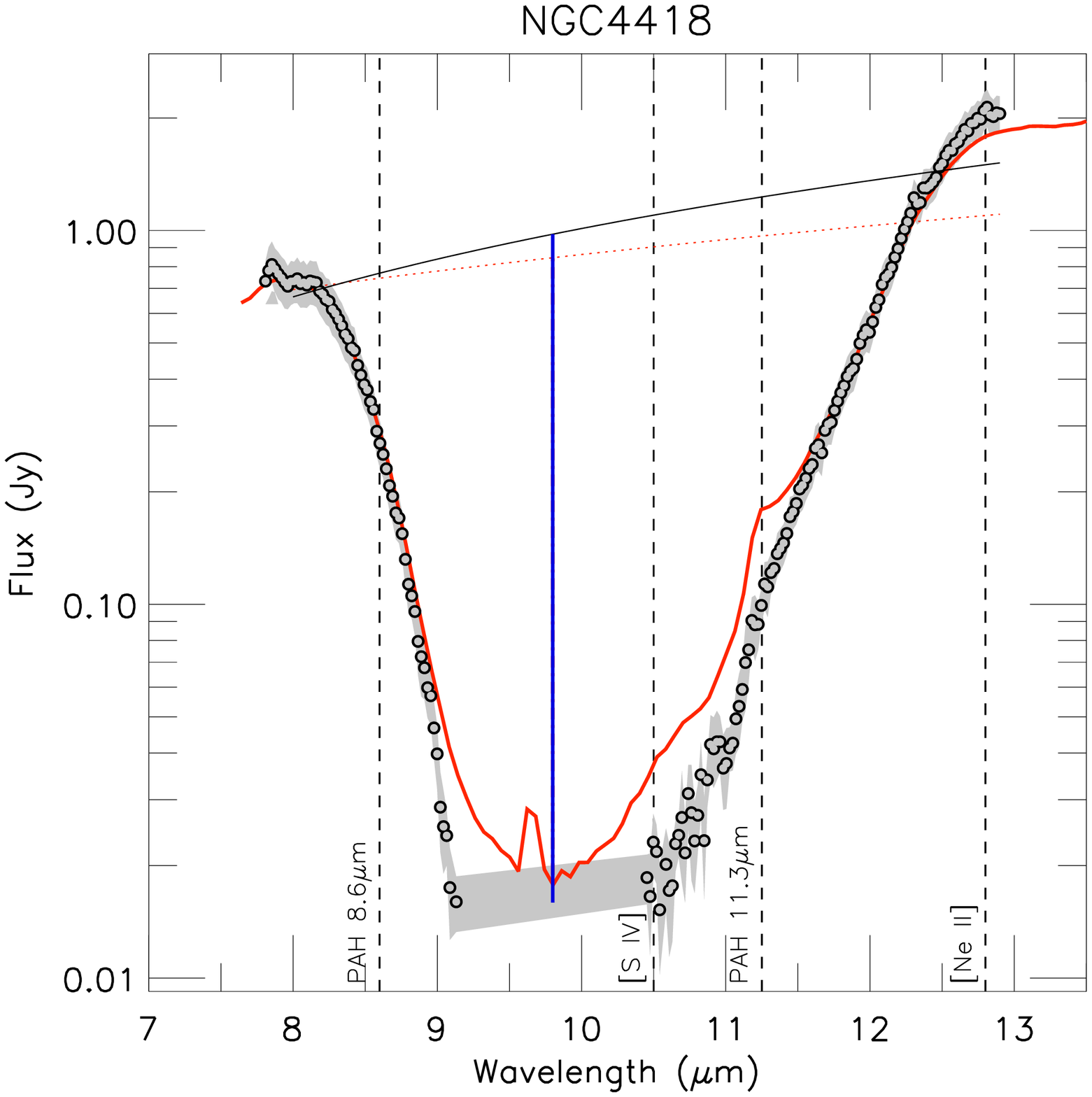}
\includegraphics[width=0.3\columnwidth]{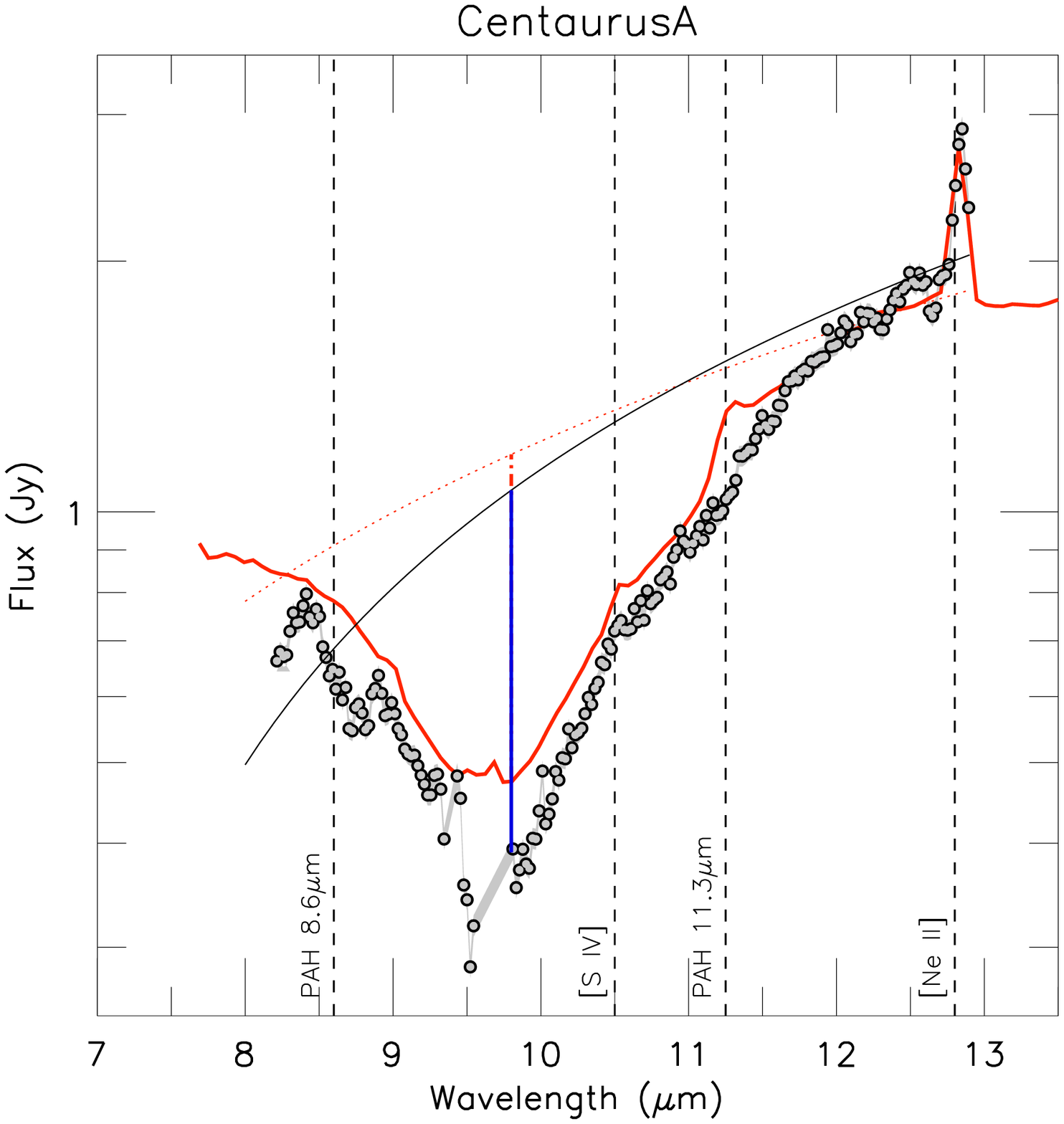}
\includegraphics[width=0.3\columnwidth]{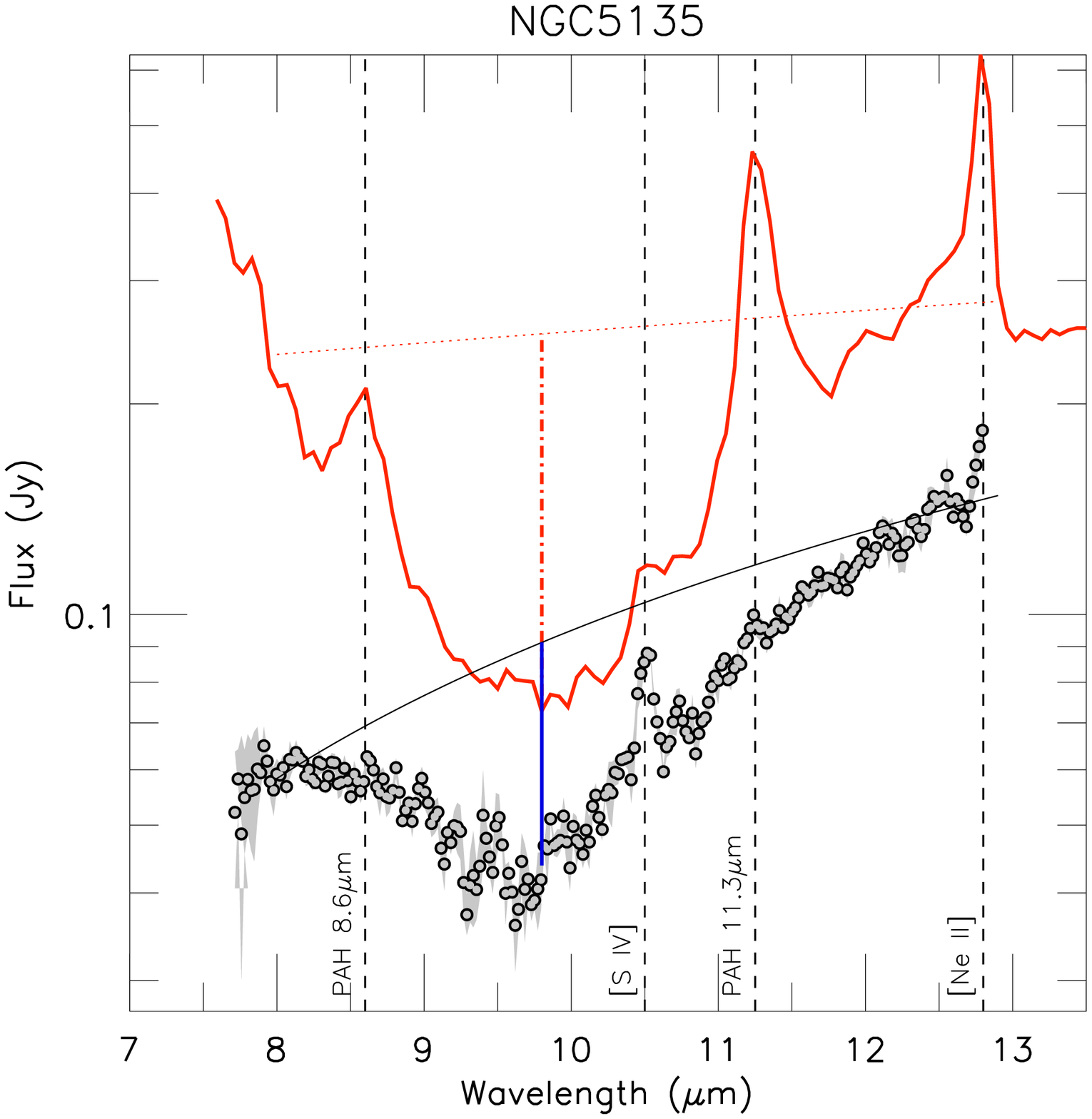}
\includegraphics[width=0.3\columnwidth]{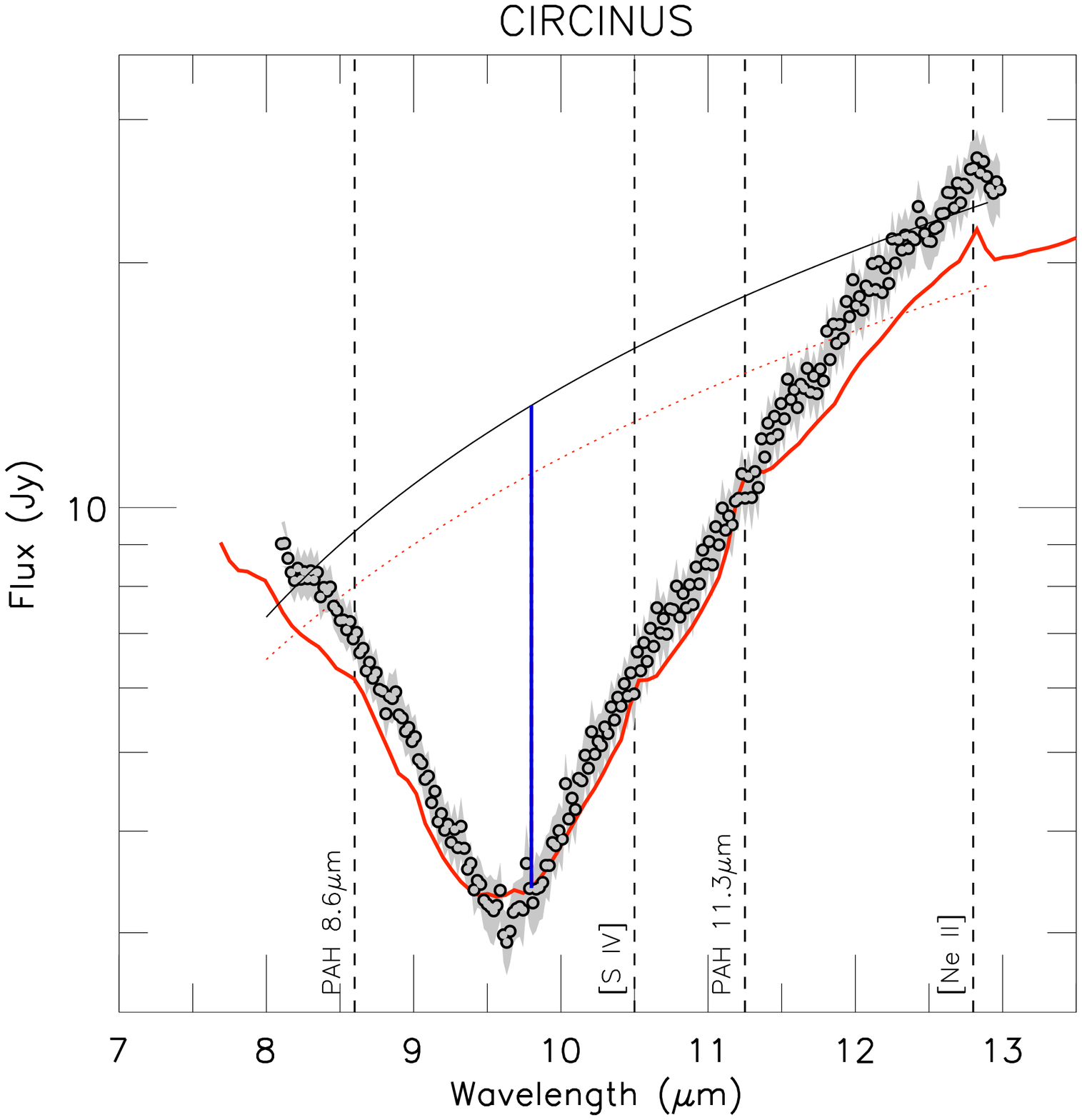}
\includegraphics[width=0.3\columnwidth]{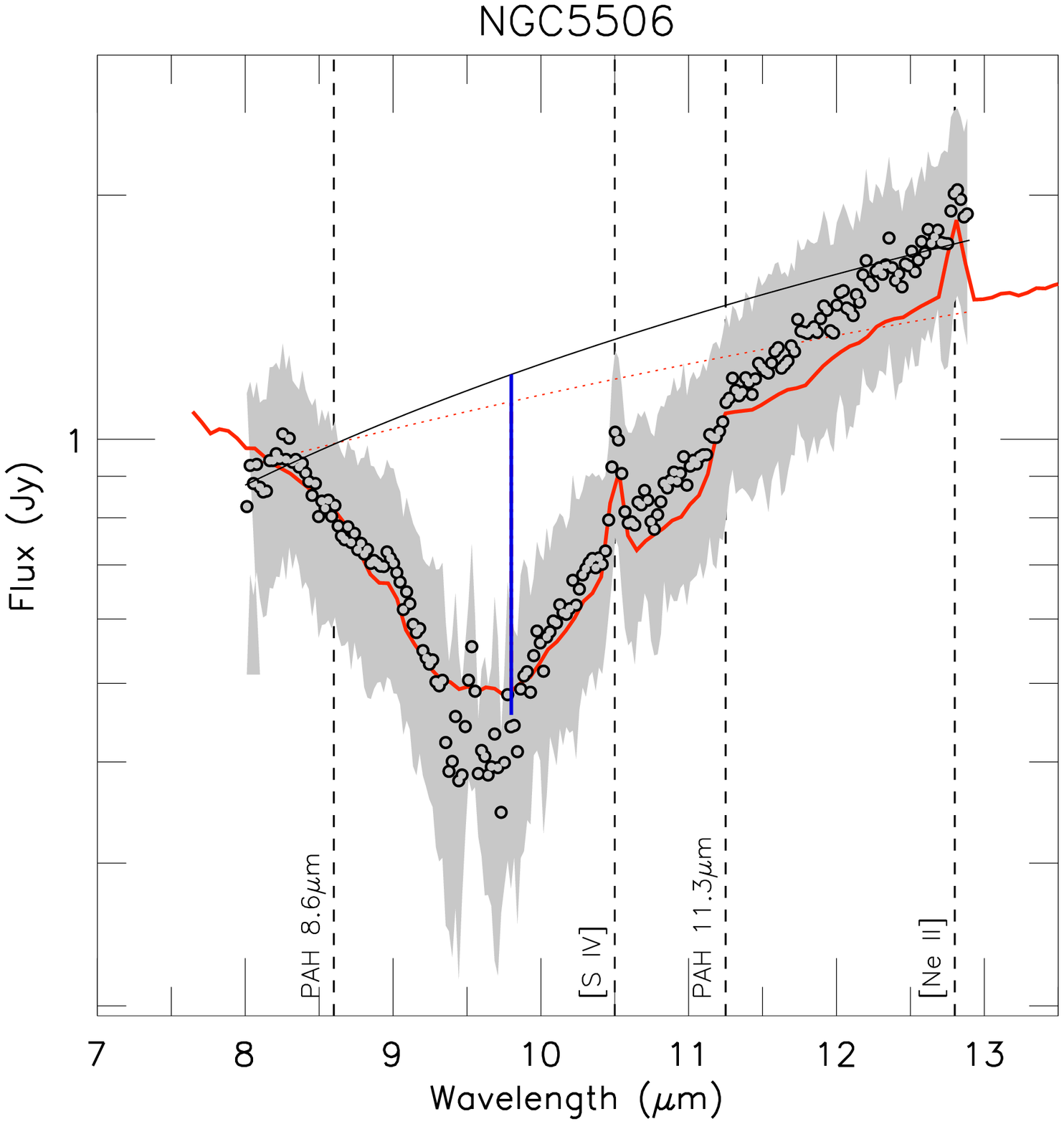}
\includegraphics[width=0.3\columnwidth]{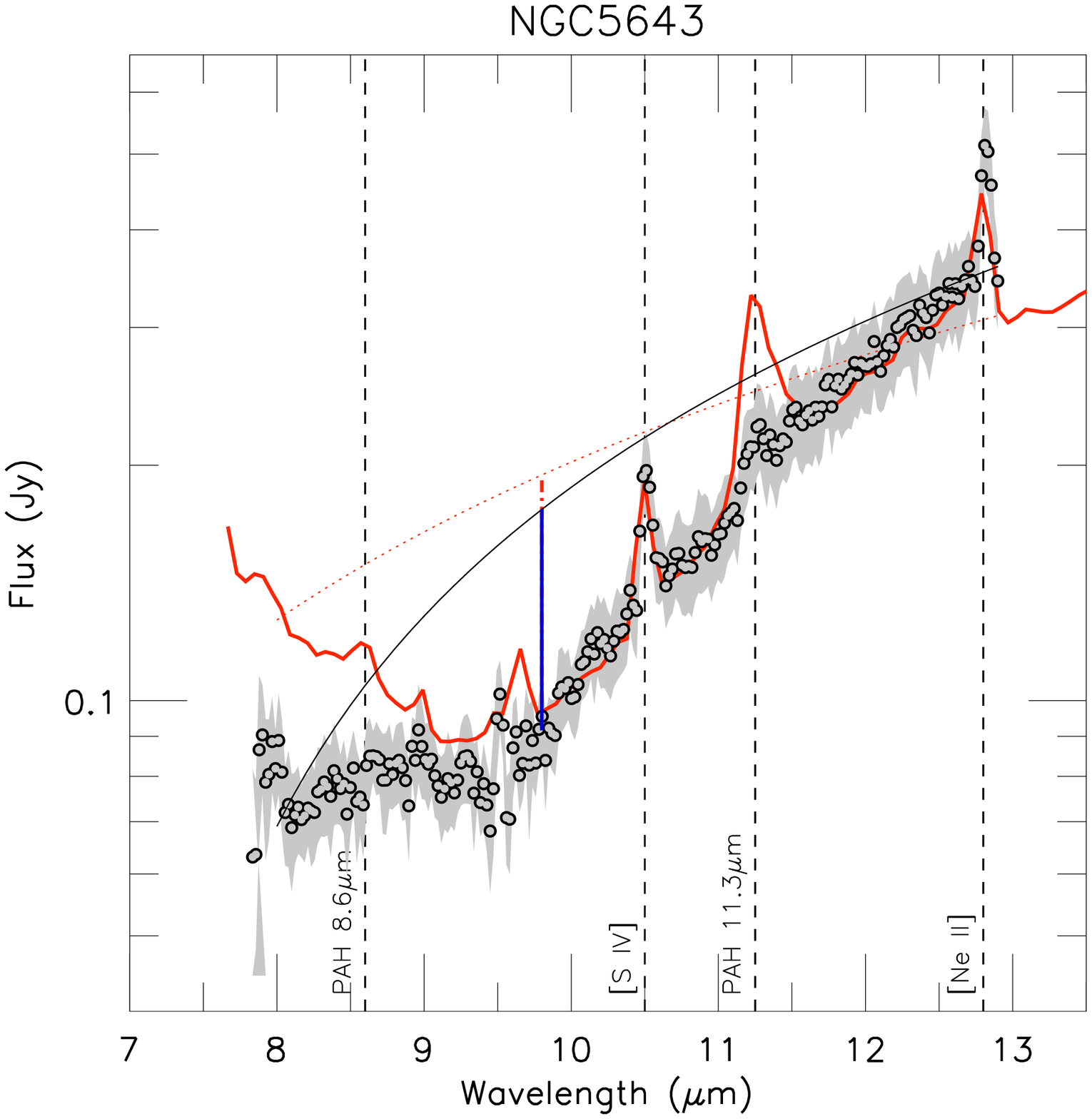}
\caption{T-ReCS spectra of the sample are shown as open circles. The \emph{Spitzer}/IRS 
spectra are the red lines. The continuum fits to the T-ReCS and IRS spectra are shown as 
solid and dotted lines, respectively. Positions of the most prominent emission lines are 
shown by dashed lines.}
\label{fig:Compilation} 
\end{center}
\end{figure}

\newpage

\begin{figure}
\begin{center}
\includegraphics[width=0.3\columnwidth]{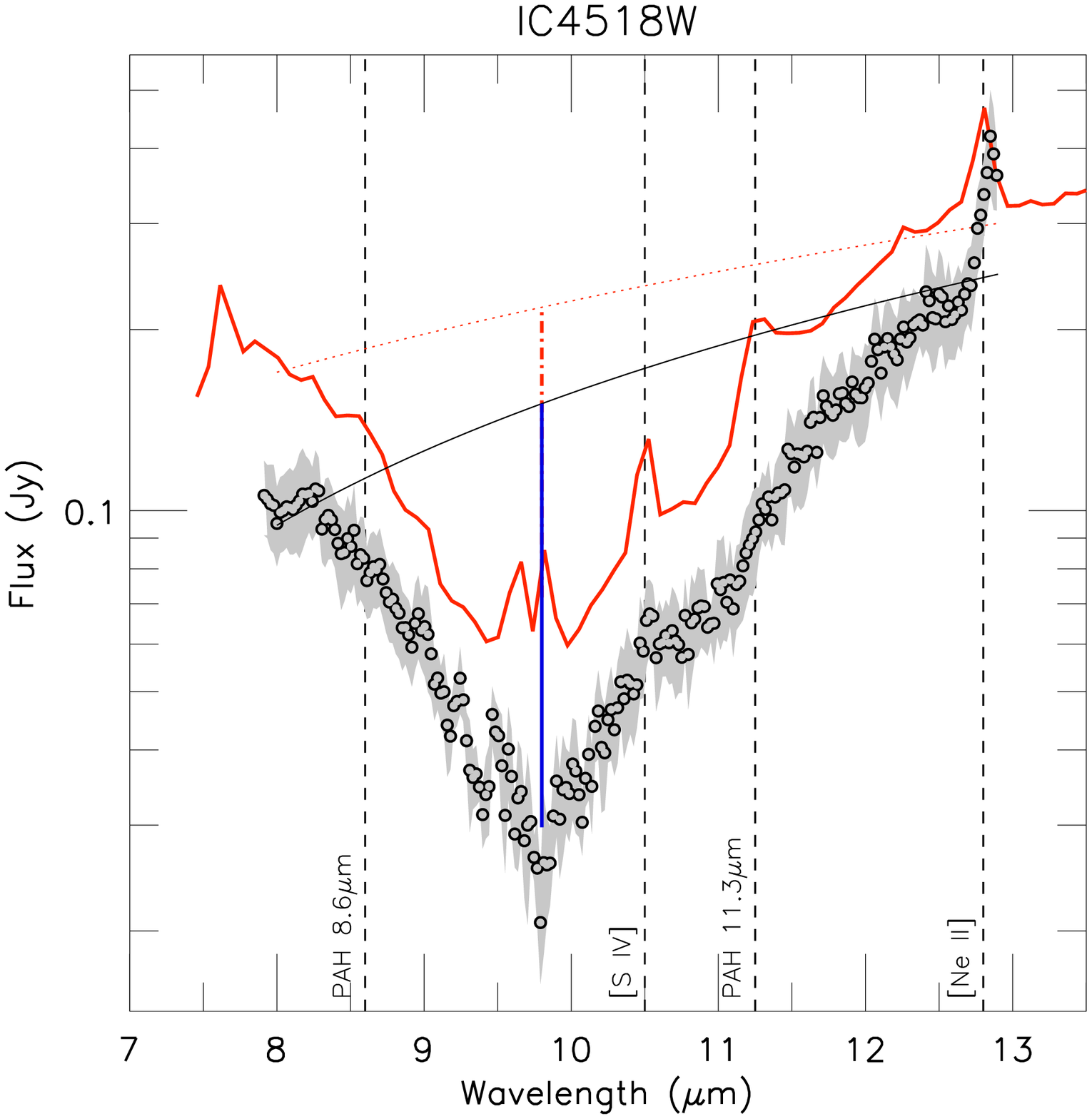}
\includegraphics[width=0.3\columnwidth]{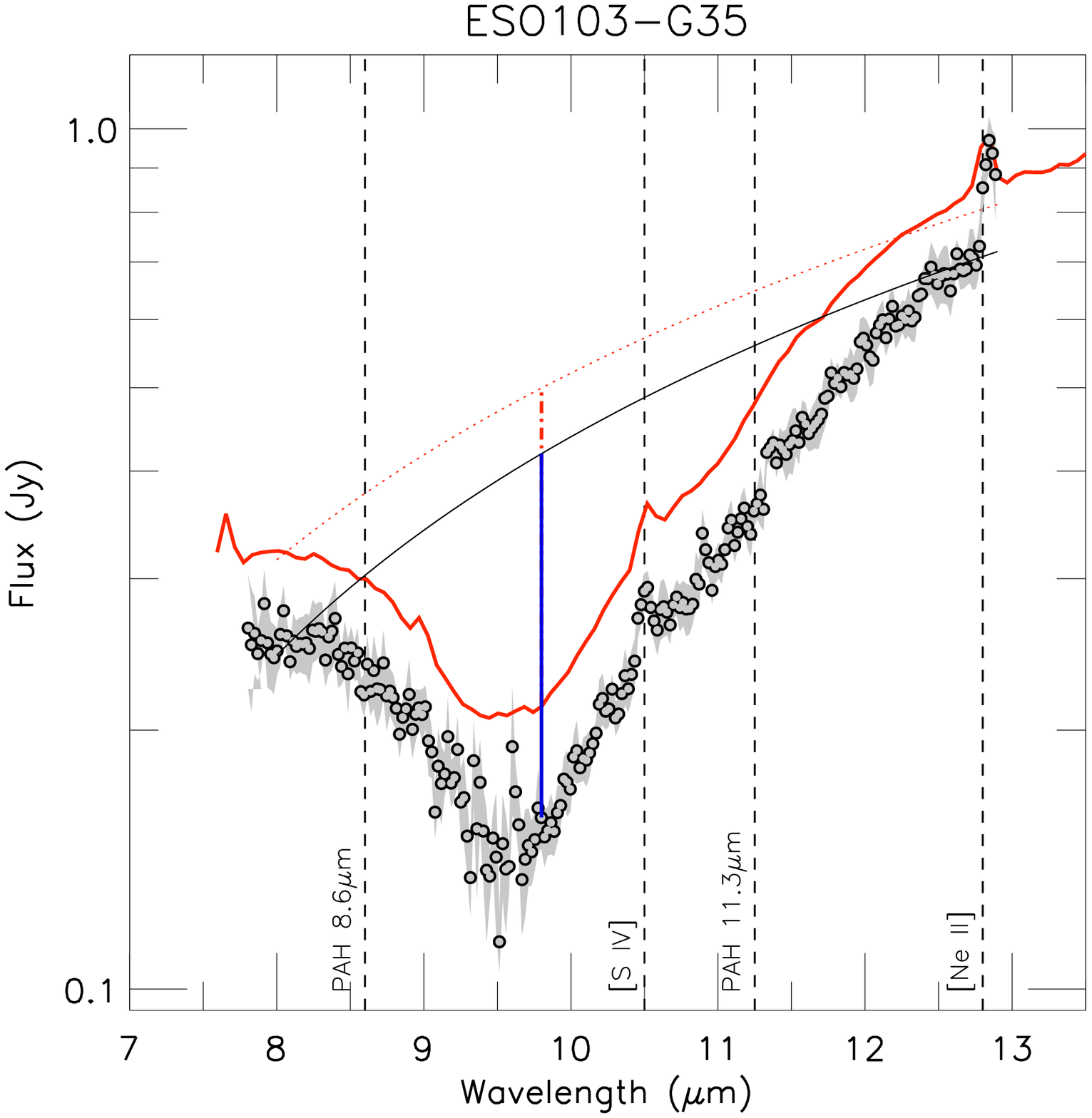}
\includegraphics[width=0.3\columnwidth]{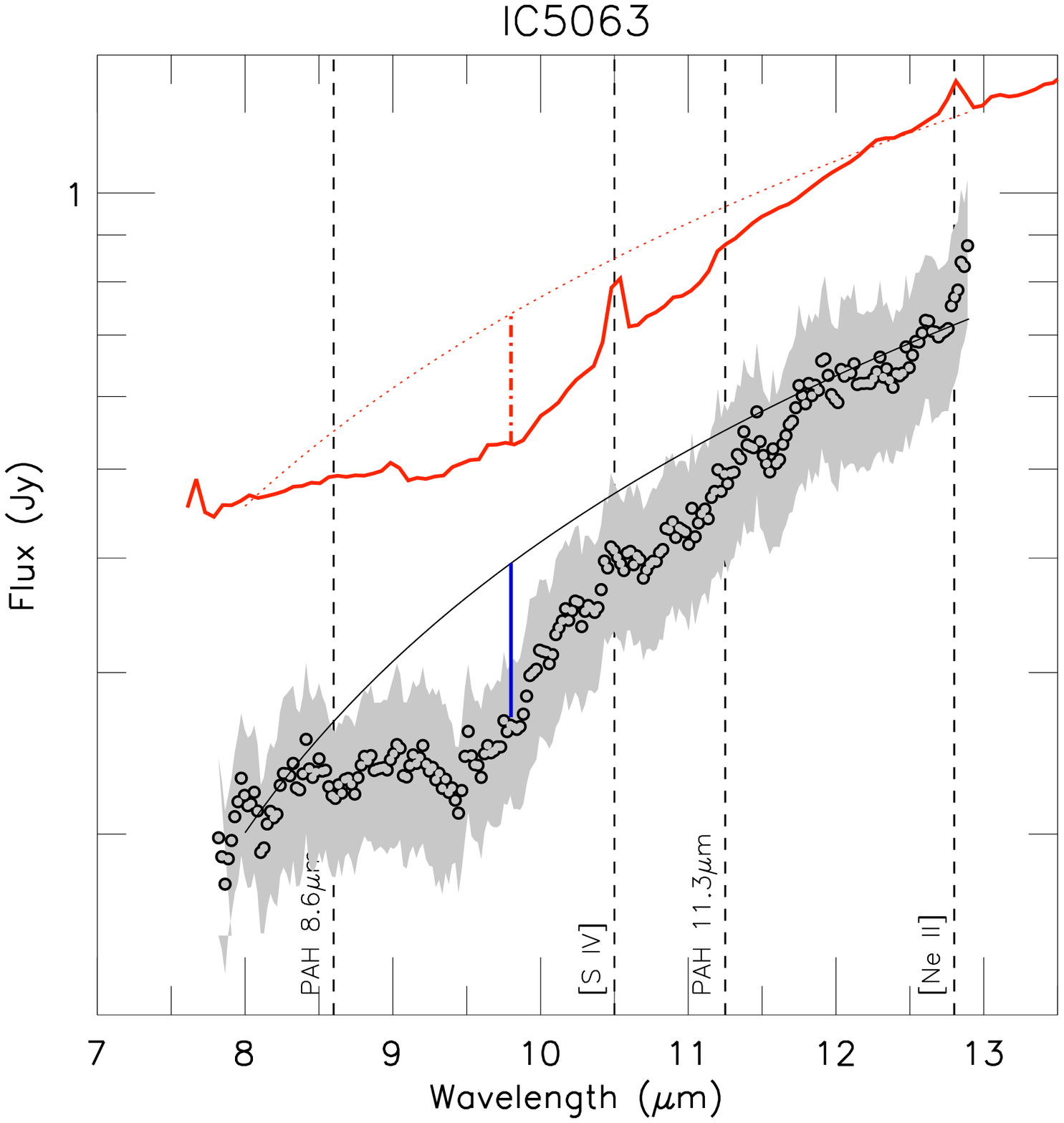}
\includegraphics[width=0.3\columnwidth]{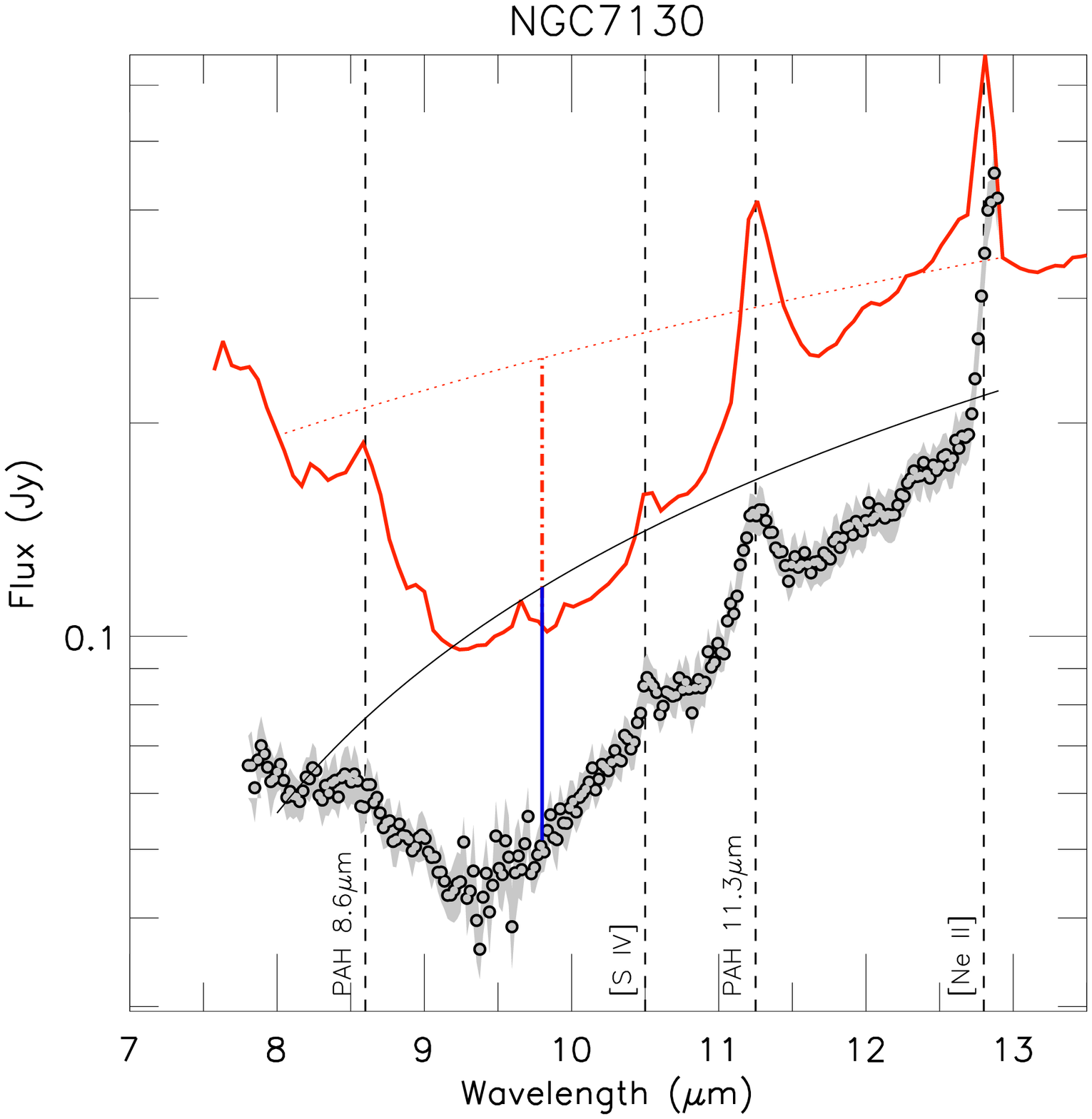}
\includegraphics[width=0.3\columnwidth]{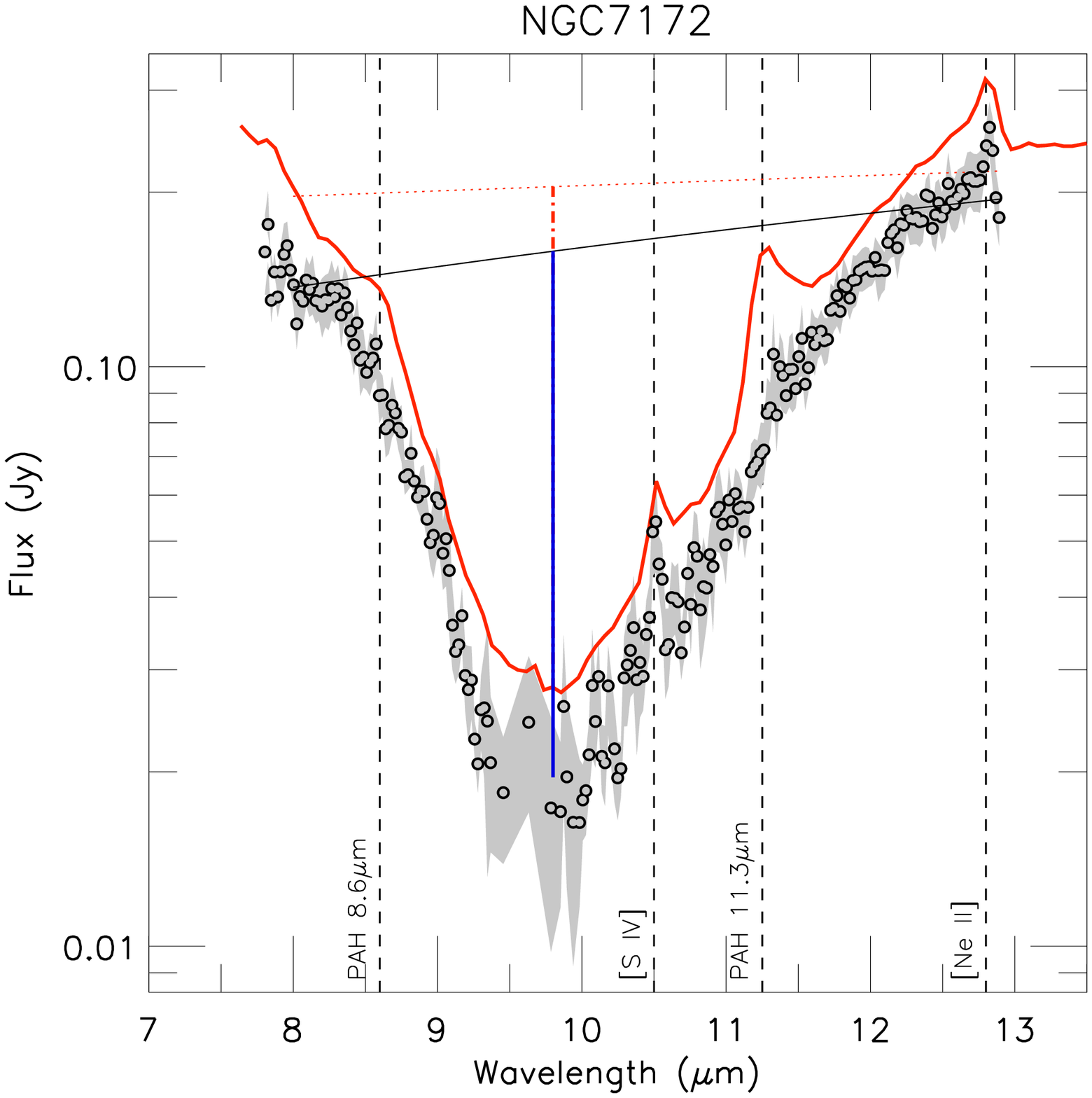}
\includegraphics[width=0.3\columnwidth]{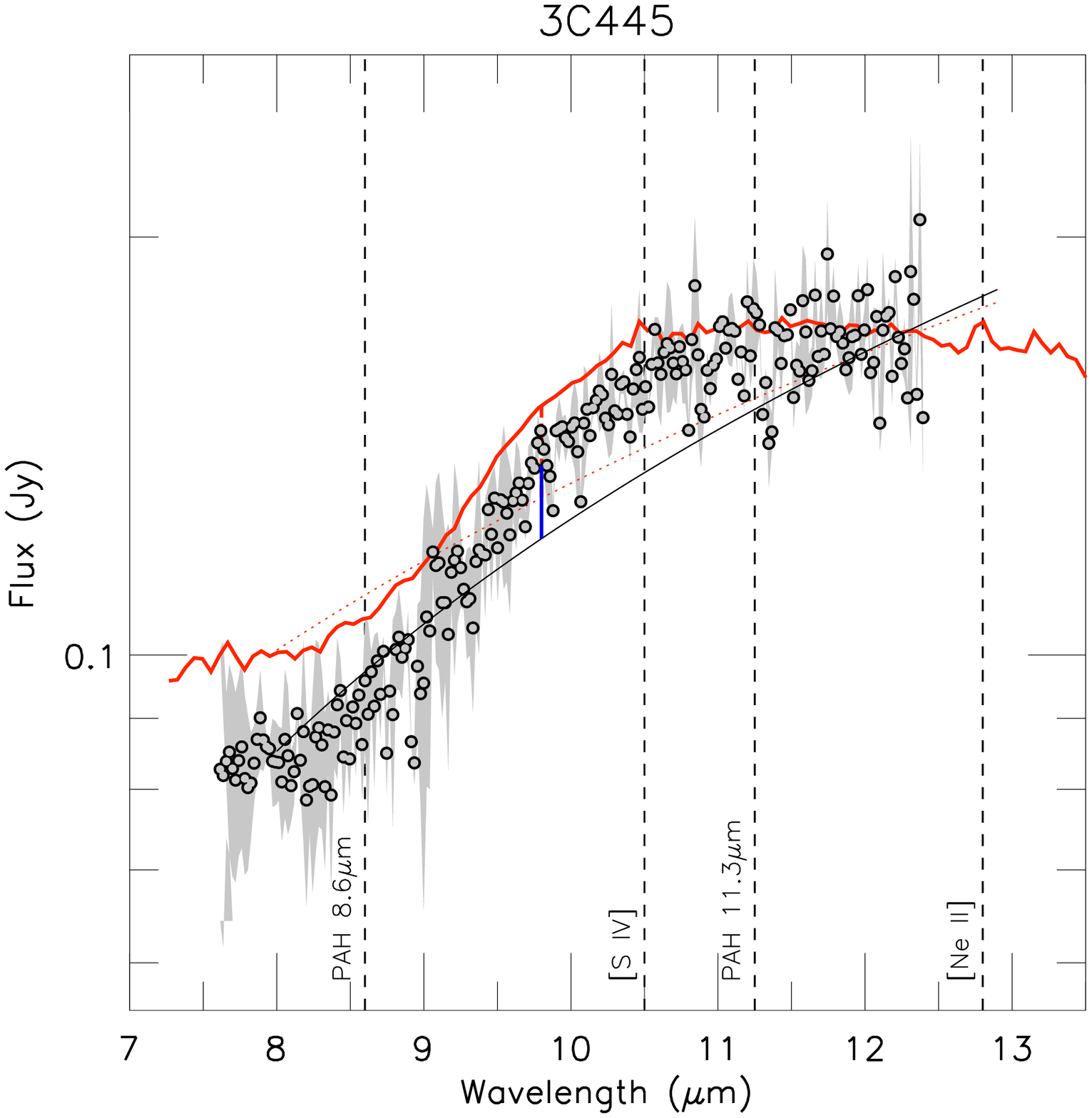}
\includegraphics[width=0.3\columnwidth]{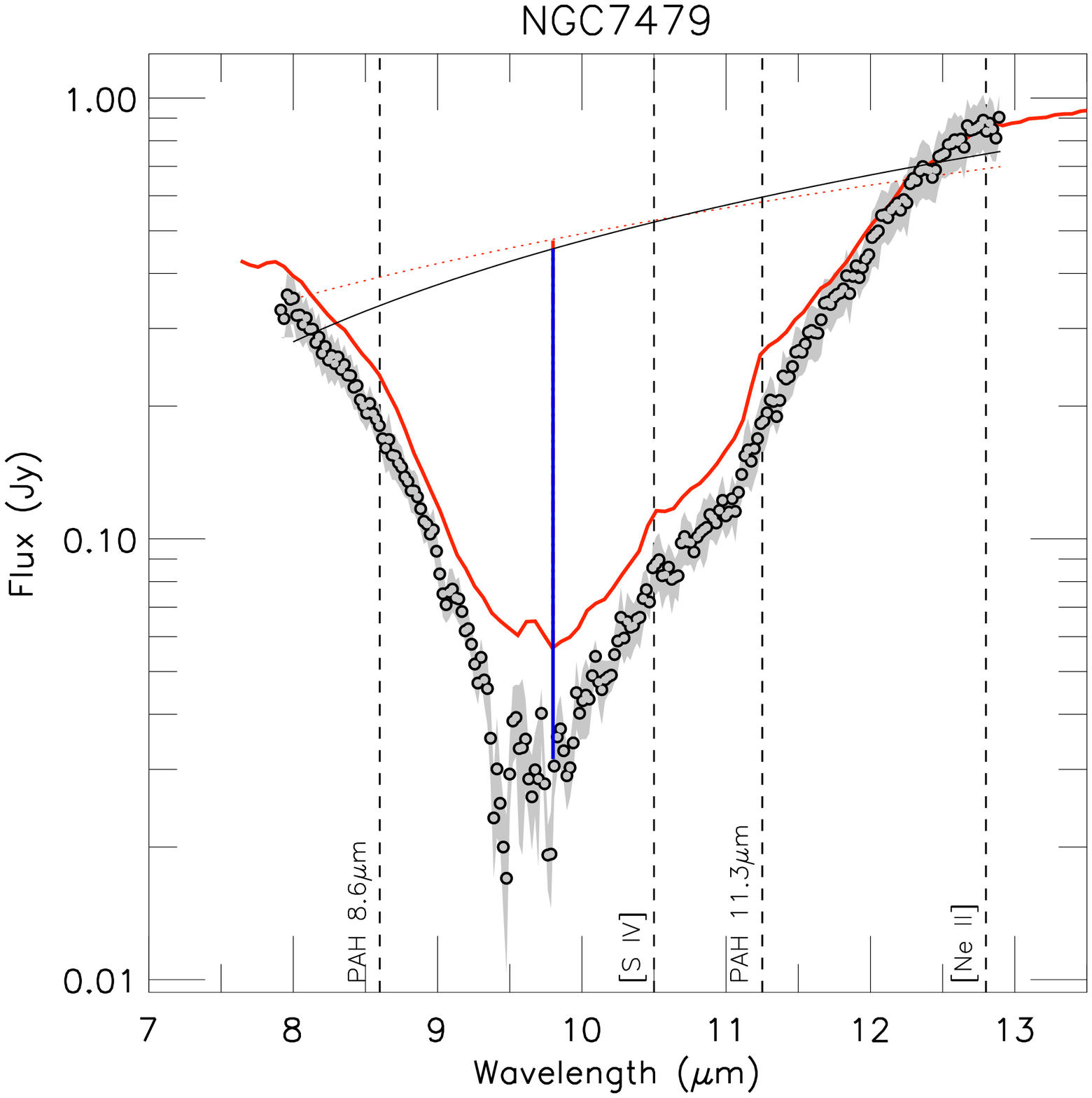}
\includegraphics[width=0.3\columnwidth]{NGC7582_spec.ps}
\setcounter{figure}{0}
\caption{T-ReCS spectra of te sample. Continued.}
\label{fig:Compilation} 
\end{center}
\end{figure}


\end{document}